\documentclass[electronics,article,accept,pdftex,moreauthors]{Definitions/mdpi}

\firstpage{1} 
\pubvolume{14}
\issuenum{1}
\articlenumber{87}
\pubyear{2025}
\copyrightyear{2024}
\externaleditor{Ping-Feng Pai}
\datereceived{20 November 2024} 
\daterevised{21 December 2024}
\dateaccepted{25 December 2024} 
\datepublished{28 December 2024} 
\hreflink{https://doi.org/10.3390/\linebreak electronics14010087} 

\Title{Cognitive Agents Powered by Large Language Models for Agile Software Project Management}

\TitleCitation{Cognitive Agents Powered by Large Language Models for Agile Software Project Management}

\Author{{Konrad} 
 Cinkusz \orcidA{}, Jarosław A. Chudziak \orcidB{} and Ewa Niewiadomska-Szynkiewicz *\orcidC{}}

\AuthorNames{Konrad Cinkusz, Jarosław A. Chudziak, and Ewa Niewiadomska-Szynkiewicz}

\AuthorCitation{{Cinkusz,} 
 K.; Chudziak, J.A.; Niewiadomska-Szynkiewicz, E.}

\address[1]{Faculty of Electronics and Information Technology, Warsaw University of Technology, 00-661 Warsaw, Poland; {konrad.cinkusz.stud@pw.edu.pl} (K.C.); {jaroslaw.chudziak@pw.edu.pl} (J.A.C.)
}

\corres{\hangafter=1 \hangindent=1.05em \hspace{-0.82em}Correspondence: {ewa.szynkiewicz@pw.edu.pl}}

\abstract{{This} 
 paper investigates the integration of cognitive agents powered by Large Language Models (LLMs) within the Scaled Agile Framework (SAFe) to reinforce software project management. By deploying virtual agents in simulated software environments, this study explores their potential to fulfill fundamental roles in IT project development, thereby optimizing project outcomes through intelligent automation. Particular emphasis is placed on the adaptability of these agents to Agile methodologies and their transformative impact on decision-making, problem-solving, and collaboration dynamics. The research leverages the CogniSim ecosystem, a platform designed to simulate real-world software engineering challenges, such as aligning technical capabilities with business objectives, managing interdependencies, and maintaining project agility. Through iterative simulations, cognitive agents demonstrate advanced capabilities in task delegation, inter-agent communication, and project lifecycle management. 
By employing natural language processing to facilitate meaningful dialogues, these agents emulate human roles and improve the efficiency and precision of Agile practices. Key findings from this investigation highlight the ability of LLM-powered cognitive agents to deliver measurable improvements in various metrics, including task completion times, quality of deliverables, and communication coherence. These agents exhibit scalability and adaptability, ensuring their applicability across diverse and complex project environments. This study underscores the potential of integrating LLM-powered agents into Agile project management frameworks as a means of advancing software engineering practices. This integration not only refines the execution of project management tasks but also sets the stage for a paradigm shift in how teams collaborate and address emerging challenges. By integrating the capabilities of artificial intelligence with the principles of Agile, the CogniSim framework establishes a foundation for more intelligent, efficient, and adaptable software development methodologies.
}


\begin{document}

\maketitle

\section{Introduction}
\label{sec:introduction}

Software engineering has undergone significant transformations due to rapid technological progress. The integration of sophisticated tools and methodologies is now essential to manage the increasing complexity and scale of modern software systems~\cite{abrahamsson_agile_2017, perkusich_intelligent_2020}.

\subsection{Background}

\textls[-15]{In this rapidly changing landscape, the adoption of advanced technologies plays a critical role in addressing the challenges posed by large-scale and complex software systems. Today's software development environment faces significant challenges, including managing extensive codebases, ensuring security, and maintaining quality across distributed teams. Traditional methodologies, such as the Waterfall model, have given way to Agile frameworks that emphasize iterative development, customer collaboration, and flexibility~\cite{dingsoyr_decade_2012}. }

Despite the advantages of Agile methodologies, they have inherent limitations when scaling and managing large-scale, complex projects effectively~\cite{shastri_role_2021}. Managing complex software systems requires breaking down the development process into structured activities. A common approach involves applying generic activities, communication, planning, modeling, construction, and deployment for each major product function~\cite{pressman_software_2020}.

\begin{figure}[H]

  \includegraphics[width=0.82\columnwidth,keepaspectratio]{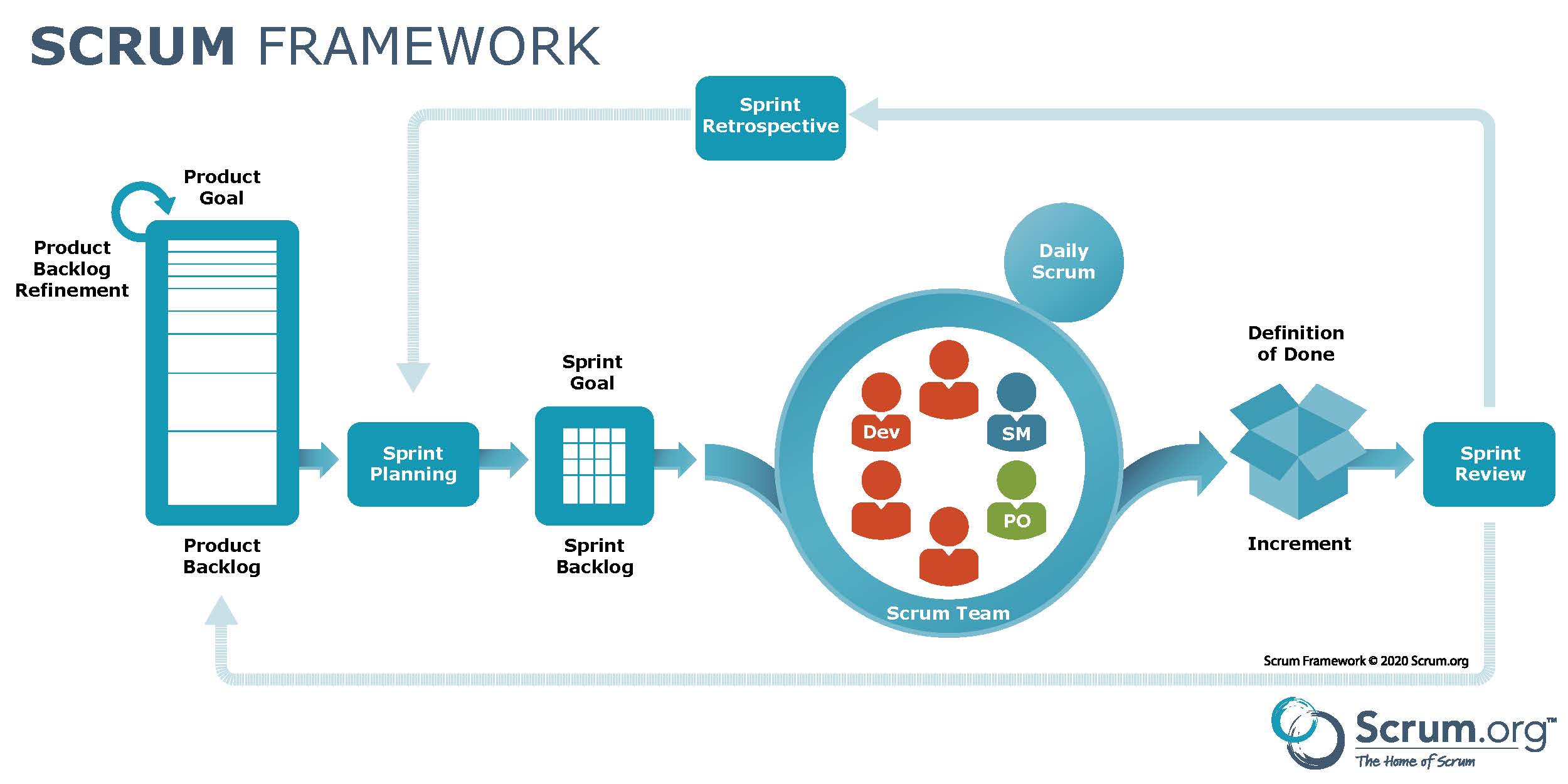}
  \caption{{Scrum} framework with key artifacts, meetings, and processes~\cite{scrumframework2020}{.} 
}
  \label{fig:SCRUM_framework}
\end{figure}

The Scrum, a framework within Agile methodologies, addresses large-scale software system challenges through structured processes and defined roles. As shown in Figure~\ref{fig:SCRUM_framework}, it organizes work via artifacts and scheduled meetings. The Product Backlog, managed by the Product Owner, aligns requirements with stakeholder goals. The Scrum team, led by the Scrum Master, plans a Sprint Backlog for a time-boxed iteration called a Sprint. Daily Scrums track progress, while the Sprint Review assesses deliverables, and the Sprint Retrospective identifies improvements ~\cite{rubin_essential_2013}. This cycle ensures transparency, accountability, and steady progress toward a quality product.
\vspace{-9pt}

\begin{figure}[H]

  \includegraphics[width=0.8\columnwidth,keepaspectratio]{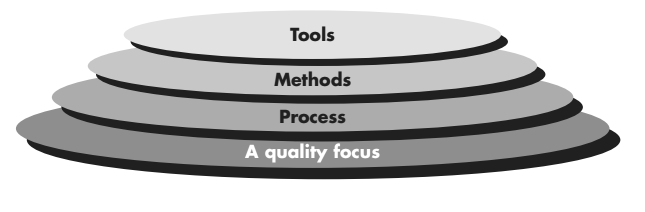}
  \caption{Software engineering layers~\cite{pressman_software_2020}.}
  \label{fig:layered_technology_framework}
\end{figure}

As detailed in Figure~\ref{fig:layered_technology_framework}, software engineering is built on layered technology where each layer contributes to the quality and structure of the development process. This layered approach encompasses a focus on quality, processes, methods, and tools, collectively facilitating the systematic management of software engineering tasks~\cite{pressman_software_2020}. It enables project teams to estimate resource requirements, schedule tasks, and define work products for each development stage.

Multi-Agent Systems, comprising networks of autonomous agents that interact to achieve defined objectives within their environment, offer a resilient solution to these challenges~\cite{cruz_transforming_2024}. These agents collaborate seamlessly, emulating human problem-solving processes and contributing to overall system efficiency~\cite{spanoudakis_engineering_2021}.

Simultaneously, Large Language Models such as OpenAI's GPT-4, {{along} 
\textls[-25]{ with others like Google’s PaLM and Meta’s LLaMA, have transformed natural language processing by generating human-like content across various formats, including natural text generation for conversational tasks and storytelling, code generation to assist in software development, and tool use to perform complex workflows such as web searching, robotic operations, and online shopping~\cite{guo_evaluating_2023}.} In software engineering, LLMs can automate routine tasks like code completion, documentation, and debugging, reducing errors and leading to higher productivity~\cite{barua_exploring_2024, dvivedi_comparative_2024, tian_debugbench_2024}.}
 
{
Scaling Agile methodologies for large, complex projects introduces challenges such as coordinating across multiple teams, maintaining communication, and aligning organizational goals. Frameworks like SAFe, LeSS, and DAD address these issues by offering structured yet flexible approaches to Agile on scale, including defined roles, processes, and lifecycle models tailored to project needs. Although adoption requires overcoming cultural resistance and integration challenges, these frameworks have shown improvements in product quality, customer satisfaction, and delivery speed \cite{tian_debugbench_2024}.
}

\subsection{Related Work}

{
Multi-Agent Systems (MASs) have been applied to distributed problems in dynamic environments \cite{yuan_intelligent_2001}, and the FIPA standards \cite{fipa2012designprocess} established by the IEEE Computer Society enhance interoperability and efficiency within Agile frameworks. These standards streamline agent communication, coordination, and scalability. When combined with LLMs, as in frameworks like CoALA, MASs benefit from improved memory management, adaptive decision-making, and fluid inter-agent cooperation \cite{sumers2024cognitive}.
}

{
Recent advancements in MASs and LLMs are reshaping software engineering by introducing cognitive capabilities that improve reasoning, planning, and teamwork \cite{cinkusz_towards_2024}. LLM-based MASs have emerged as a promising vision, equipped to address complex engineering tasks through autonomous problem-solving and scalable coordination~\cite{he_llm-based_2024}. Frameworks such as AGILE integrate reinforcement learning and LLMs, enabling agents to leverage tools and consult domain experts more effectively~\cite{feng_agile_2024}. These agents have demonstrated applicability throughout the software lifecycle—from for requirements from 
engineering to maintenance—supported by specialized benchmarks and evaluation methods \cite{jin_llms_2024}. CodePori exemplifies the scalable integration of LLM-MAS, streamlining development tasks such as design, coding, and testing~\cite{rasheed_codepori_2024}.
}

{
Over the past decade, Agile methodologies have significantly matured, with research underscoring the importance of theory-driven approaches to enhance their effectiveness \cite{dingsoyr_decade_2012}. Simultaneously, incorporating intelligent techniques, including machine learning and Bayesian networks, into Agile software development (ASD) has been shown to support decision-making, optimize effort estimation, and refine resource allocation \cite{perkusich_intelligent_2020}. Advances in MASs combined with LLMs illustrate how LLM-powered agents can promote more nuanced negotiation, role specialization, and collaborative decision-making \cite{talebirad_multi-agent_2023}. Such systems can shift organizational dynamics from competition toward cooperative efforts, improving knowledge exchange and communication \cite{cruz_transforming_2024}. Furthermore, integrating theory-of-mind capabilities into LLM-based MASs promises richer context-sensitive interactions and deeper inter-agent comprehension \cite{li_theory_2023}. 
}

\subsection{Motivation and Research Gap}

The combination of Multi-Agent Systems and Large Language Models creates a significant synergy, giving rise to cognitive multi-agent ecosystems that merge the strengths of both technologies~\cite{li_theory_2023, singhal_publisher_2023}. This integration is particularly relevant in Agile software development, where flexibility, collaboration, and customer-centric approaches are paramount~\cite{abrahamsson_agile_2017}. LLM-augmented MASs can facilitate more efficient task performance, adapting to dynamic contexts and evolving requirements, and thereby more effectively address the complexities inherent in modern large-scale software engineering projects.

{
At the same time, frameworks for scaling Agile practices—such as the Scaled Agile Framework, Large-Scale Scrum (LeSS), or Disciplined Agile Delivery (DAD)—provide structured guidance for extending Agile principles across diverse and distributed teams~\cite{safe6.0, shastri_role_2021}, while these frameworks have achieved measurable improvements in areas like product quality and delivery speed, they still struggle to fully accommodate the heightened complexity and coordination demands of large-scale software initiatives. Existing approaches often lack the cognitive sophistication required to integrate advanced decision-making support and context-aware communication, leaving critical opportunities for more robust reasoning, planning, and adaptability unaddressed.
}

{
Although MASs have shown potential in distributed problem-solving and adaptive coordination~\cite{fipa2012designprocess, yuan_intelligent_2001}, and LLMs have demonstrated effectiveness in tasks such as code generation, documentation, and interactive tool use~\cite{guo_evaluating_2023, barua_exploring_2024, dvivedi_comparative_2024, tian_debugbench_2024}, the literature lacks a cohesive framework that unifies these capabilities within a scaled Agile context. Recent work has explored cognitive MASs and LLMs individually~\cite{sumers2024cognitive, cinkusz_towards_2024, he_llm-based_2024, feng_agile_2024}, yet the integration of these technologies into established scaled Agile processes remains limited and fragmented.
}

{
Integrating cognitive agents and LLMs into frameworks like SAFe could provide deep, context-sensitive insights into project management and development workflows, enhancing not only the efficiency and quality of engineering tasks but also the agility with which teams respond to evolving objectives~\cite{kim_financial_2024, chiang_can_2023}. By explicitly addressing the current research gap—namely, the absence of a comprehensive, LLM-augmented MAS framework aligned with scaled Agile methodologies—this work aims to establish a foundation that can lead to improved coordination, decision-making, and adaptability in large-scale software engineering efforts.
}

\subsection{Objectives and Problem Statements}
{
This study aims to develop and analyze the CogniSim framework, a cognitive Multi-Agent System designed to transform software project management by integrating cognitive agents powered by LLMs. The primary objectives are to create a framework that automates routine project tasks, enhances workflows, and aligns with established Agile practices—particularly SAFe—to ensure scalability and effectiveness. By demonstrating its practical applications in software engineering, the framework will be evaluated through case studies and simulations.
}

{
Building upon the identified research gap and the outlined objectives for integrating LLM-augmented MASs within Scaled Agile Frameworks, this study’s evaluation focuses on the following research questions (RQs):
\begin{enumerate}
    \item {{RQ1: To what extent can cognitive agents, powered by LLMs, effectively simulate Agile roles and processes in a complex software development environment (e.g., SAFe)?} 
}
    \item {{RQ2: How do variations in key parameters (e.g., model type, number of iterations, agent roles) influence the quality of outcomes, including code artifacts, documentation, and decision-making efficacy?}}
\end{enumerate}
}

\subsection{Approach and Methodology}

{
This study adopts a design and implementation process to develop and refine the CogniSim framework as a problem-solving artifact. The methodology is aligned with Agile principles and SAFe guidelines, ensuring that the virtual environment and agent interactions reflect the complexities of large-scale software development~\cite{dingsoyr_decade_2012, shastri_role_2021, safe6.0}. By grounding the approach in Agile concepts and considering the evolving nature of software architectures, we address the need for reconciling architectural documentation with Agile methodologies for better project outcomes~\cite{rocha_guidelines_2023}.
}

{
\textls[-25]{Each cognitive agent represents a distinct Agile role—such as a Product Owner, System Architect, or QA Engineer—and operates autonomously to manage tasks, perform quality assurance, and provide continuous feedback. By doing so, the system emulates typical workflows, dependencies, and communication patterns in modern software engineering projects~\cite{abrahamsson_agile_2017, perkusich_intelligent_2020, pressman_software_2020}.}
}

{
The technical foundation involves Python-based tooling, GPT-4 and GPT-3.5 language models, and the LangChain framework. LLM-powered agents process natural language instructions, generate documentation, and make context-aware decisions. This capacity allows the agents to adapt to evolving project goals, resource constraints, and shifts in priority, supporting Agile methodologies that emphasize responsiveness to change~\cite{barua_exploring_2024, dvivedi_comparative_2024, tian_debugbench_2024}.
}

{
This methodology uses iterative simulations and case studies to evaluate the system’s performance under various conditions. Simulated scenarios include activities such as Program Increment (PI) Planning, Iteration Execution, and Inspect-and-Adapt Workshops, all of which are core elements of SAFe-based Agile processes. These simulations enable controlled experimentation with different agent configurations, memory management strategies, and prompt designs to assess the system’s effectiveness, efficiency, and scalability~\cite{cinkusz_towards_2024, he_llm-based_2024, feng_agile_2024}.
}

{
Evaluations rely on both qualitative and quantitative metrics. Key measures include the clarity and accuracy of generated artifacts, the extent to which decisions align with the defined architectural and business objectives, and the timeliness of deliverables. By comparing these outcomes against accepted best practices or human-generated baselines, the study identifies areas for improvement and informs subsequent adjustments to agent behavior, communication protocols, and underlying language model configurations~\cite{perkusich_intelligent_2020, talebirad_multi-agent_2023}. This iterative feedback loop ensures that CogniSim remains adaptable as project complexity or organizational needs evolve~\cite{li_theory_2023, singhal_publisher_2023}.
}

\subsection{Evaluation Framework}

{
The evaluation applies both qualitative and quantitative techniques to assess how well the CogniSim framework supports Agile project management practices. Each simulation run produces a detailed record of interactions, decisions, and generated artifacts, as well as configuration parameters and anomalies encountered. This comprehensive documentation enables reproducibility and clear analysis pathways.
}

{
A qualitative thematic analysis is performed on the collected data to identify patterns that align with Agile principles, including communication efficiency, iterative refinement of tasks, and conformity with SAFe guidelines. Emergent themes are contrasted against established best practices to ensure that observed improvements are not coincidental.
}
{
Quantitative metrics complement these qualitative insights, examining factors such as task completion times, clarity of generated outputs, and responsiveness to evolving requirements. These metrics provide a multifaceted perspective on performance. Although initial results suggest positive potential, future work involves extending the range of conditions tested, introducing more complex scenarios, and incorporating benchmarking against human-driven baselines to further refine the framework’s utility and reliability.
}
\subsection{An Outline of This Study}

This study begins with Section \ref{sec:introduction}, which frames the challenges of modern software engineering, emphasizing the necessity for scalable Agile practices and introducing Multi-Agent Systems and Large Language Models as transformative solutions. Next, Section \ref{sec:preliminaries} provides the theoretical foundation, covering Agile methodologies, cognitive agents, LLMs, and MAS concepts. Section \ref{sec:cognisim} details its layered architecture, integrating LLM-powered agents into Agile workflows. In the next Section \ref{sec4}, we illustrate its application through simulations, showcasing improvements in project communication, decision-making, and task execution. Then Section \ref{sec5} discusses implementation specifics, highlighting modular design and Python-based tools for adaptability. It is followed by Section \ref{sec:case_study}, which explores agent initialization, interaction protocols, and reproducibility in performance assessments. In Section \ref{sec:experiments_results}, the framework's effectiveness is validated through metrics like task efficiency, deliverable quality, and Agile adaptability. Section \ref{sec:future_work} identifies opportunities for scaling, improving human--AI collaboration, and addressing ethical challenges. Finally, Section \ref{sec9} underscores the CogniSim framework's contribution to advancing software engineering by merging LLM-powered agents with Agile methodologies.

\section{{Preliminaries} 
}
\label{sec:preliminaries}

This section provides an overview of the fundamental concepts essential to this study, including {agile software} 
 development methodologies with a focus on the Scaled Agile Framework, an introduction to cognitive agents and Large Language Models, and the fundamentals of Multi-Agent Systems and their applications in software engineering.

\subsection{Agile Software Development}

Agile software development is a group of methodologies that promote development, collaboration, and adaptability throughout the software development life cycle. Unlike the Waterfall model, Agile emphasizes customer collaboration, responses to change, and progressive delivery of tangible software~\cite{highsmith_agile_2010}.

\begin{figure}[H]

  \includegraphics[width=0.9\columnwidth,keepaspectratio]{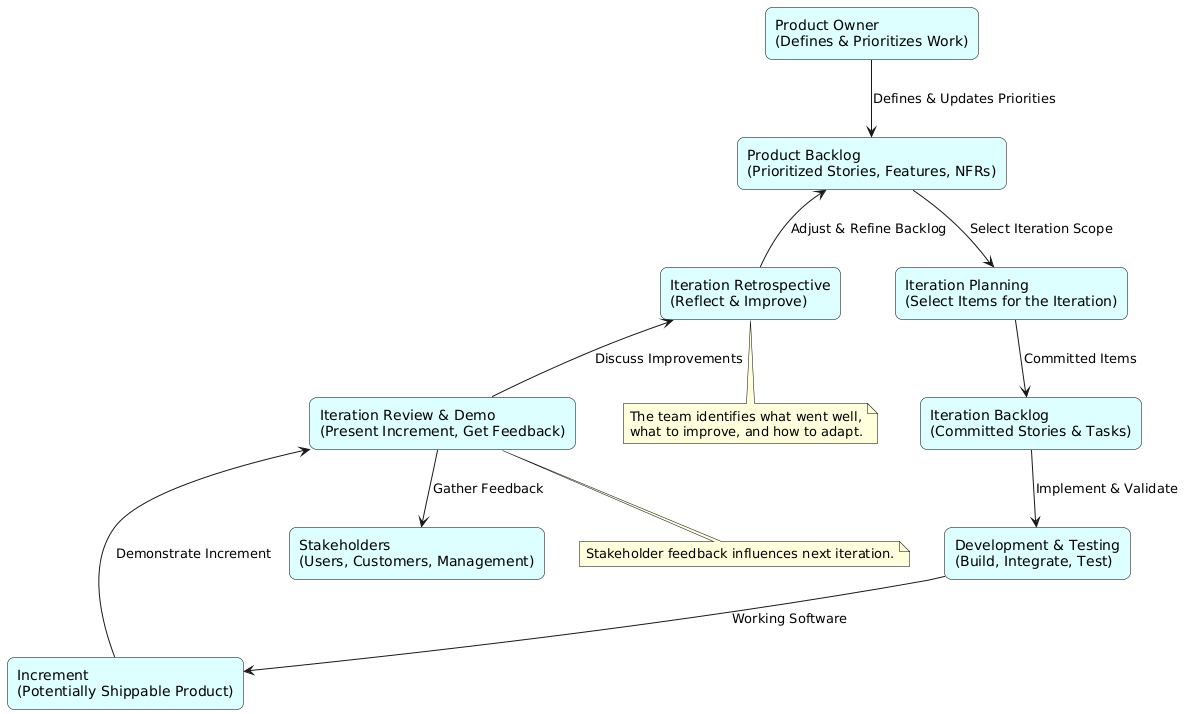}

  \caption{A generic agile iteration cycle illustrating planning, development, review, stakeholder feedback, and continuous improvement.}
  \label{fig:generic_agile_iteration}

\end{figure}

One of the most adopted agile methodologies is Scrum, which structures development into time-boxed iterations called sprints. Each sprint results in a shippable product increment, allowing for consistent reassessment of project priorities and alignment with customer needs~\cite{Schwaber2020}. As illustrated in Figure~\ref{fig:generic_agile_iteration}, a generic Agile iteration cycle typically begins by selecting items from a prioritized backlog, proceeding through development and testing, demonstrating increments to stakeholders, and reflecting in retrospectives to guide continuous improvement.

For expanding organizations dealing with advanced systems and multiple teams, the Scaled Agile Framework provides a coherent approach to scaling Agile practices~\cite{safe6.0}. SAFe integrates principles from Lean thinking, Agile development, and systems thinking to facilitate coordination between teams and align development efforts with the goals of the organization. As illustrated in Section \ref{sec:future_work}, 
the Scaled Agile Framework emphasizes five conceptual layers, Organizational Agility, Lean Portfolio Management, Enterprise Solution Delivery, Agile Product Delivery, and Team and Technical Agility, which collectively enable organizations to scale Agile practices and achieve business agility, with cognitive agents and MASs supporting each layer.

\begin{figure}[H]
  \includegraphics[width=.99\columnwidth,keepaspectratio]{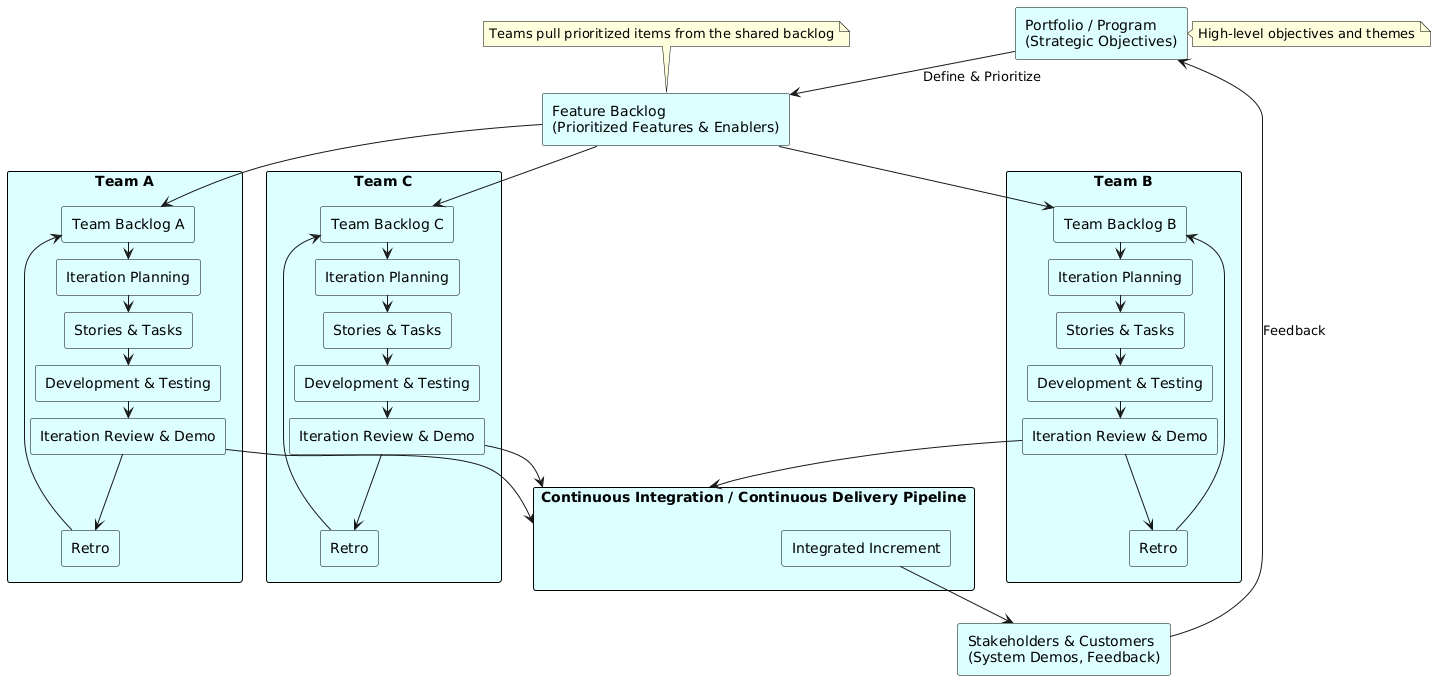}
  \caption{Conceptual scaled Agile iteration flow: multiple teams coordinating increments, integrating continuously, and aligning with strategic objectives.}
  \label{fig:conceptual_scaled_agile}

\end{figure}

\textls[-15]{SAFe introduces the concept of the Agile Release Train (ART)~\cite{knaster_safe_2020}, a structure that aligns multiple Agile teams and stakeholders to incrementally develop and deliver value within a value stream. While the ART is a SAFe-specific concept, the notion of multiple teams working in parallel and integrating their increments continuously is not exclusive to SAFe. Figure \ref{fig:conceptual_scaled_agile} offers a framework-neutral conceptual representation of how multiple Agile teams can plan, deliver, and integrate increments to support complex product development. This alignment facilitates improved coordination, synchronization, and collaboration between teams, enhancing the ability to handle advanced projects and product development.}
\vspace{-3pt}

Despite its benefits, the implementation of scaled Agile practices, including SAFe, presents challenges such as maintaining alignment between multiple teams, ensuring consistent communication, and integrating advanced technologies~\cite{ebert_scaling_2017}. The integration of pioneering technologies such as cognitive agents and LLM can address these challenges by automating coordination tasks and enhancing communication efficiency.

\subsection{Cognitive Agents and Large Language Models}

\begin{figure}[H]
  \includegraphics[width=0.6\columnwidth,keepaspectratio]{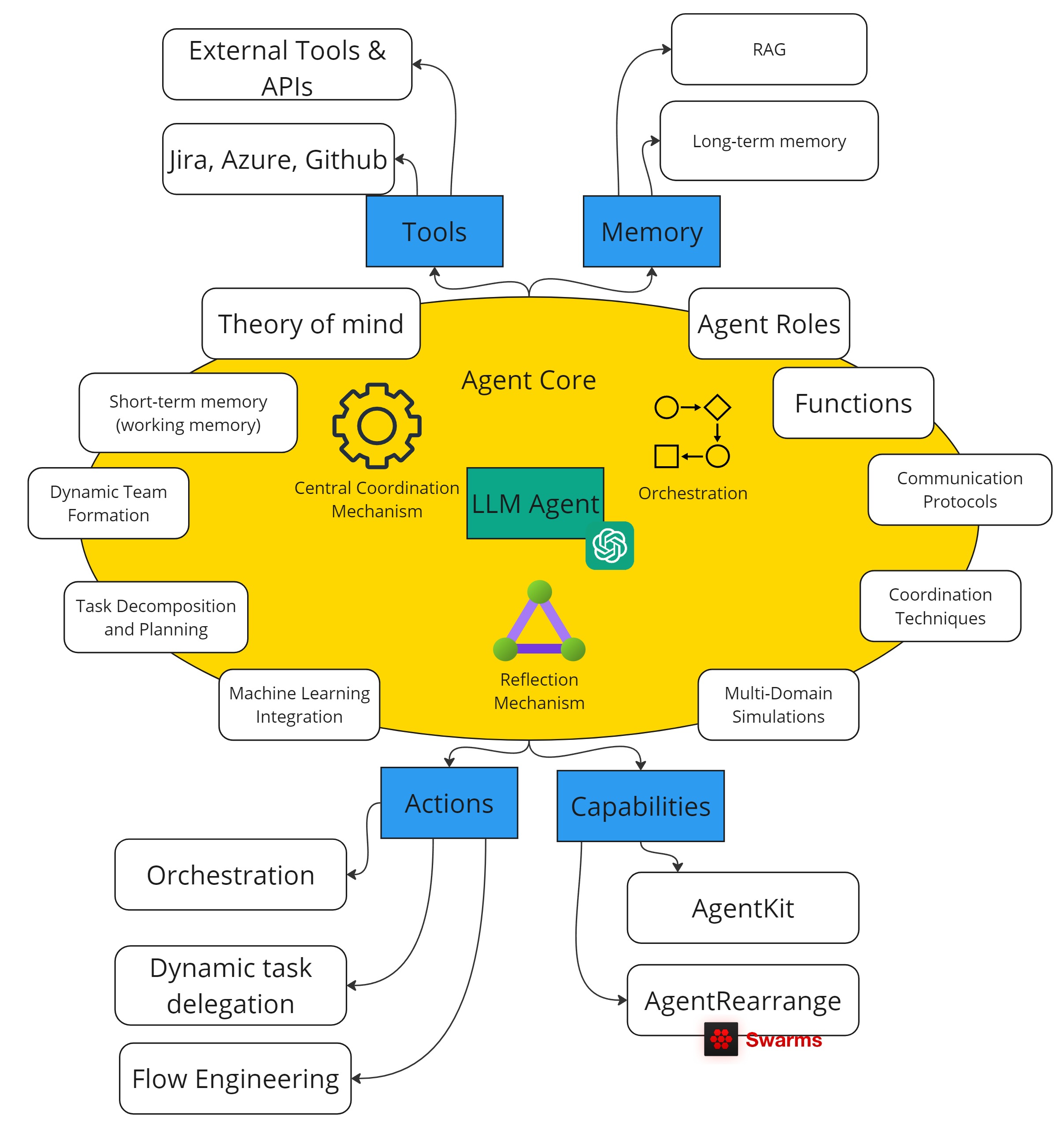}
  \caption{Single cognitive agent and its components \cite{cinkusz_towards_2024}.}
  \label{fig:single_cognitive_Agent}
\end{figure}

Cognitive agents are smart systems capable of perceiving their environment, reasoning about inputs, learning from experiences, and taking actions to achieve goals~\cite{russell_artificial_2021}. They mimic human cognitive functions such as perception, learning, and problem-solving, enabling them to perform advanced tasks autonomously. A single cognitive agent, as shown in Figure~\ref{fig:single_cognitive_Agent}, can range from a simple unit equipped with a specific function to a highly capable entity integrating diverse components such as memory, tools, and reasoning mechanisms. This flexibility allows cognitive agents to adapt to varying requirements, enabling modular scalability and dynamic task allocation based on evolving needs.

The development of Large Language Models, such as GPT-4, has advanced the capabilities of cognitive agents~\cite{openai_gpt-4_2024}. {LLMs, such as GPT-4, are trained on diverse and extensive datasets, including sources like web pages, books, and scientific articles, enabling them to understand and generate language effectively \cite{openai_gpt-4_2024}.} This capability allows cognitive agents to process input from natural language, participate in conversations, and perform tasks that require understanding context and semantics~\cite{brown_language_2020}.
\vspace{-3pt}

In the context of software engineering, LLMs enhance cognitive agents by enabling them to automate code generation~\cite{chen_evaluating_2021}, assist in documentation~\cite{svyatkovskiy_intellicode_2020}, facilitate communication~\cite{li_theory_2023}, and support decision-making~\cite{singhal_publisher_2023}. By automating code generation, cognitive agents can produce code snippets or entire functions based on natural language descriptions, streamlining the development process. Assisting in documentation involves creating or updating documentation by summarizing code functionalities, ensuring that project documentation remains updated and substantial. Facilitating communication allows these agents to serve as virtual assistants in meetings, transcribing discussions and highlighting action items, thereby improving team collaboration. Additionally, supporting decision-making enables agents to analyze project data and provide insights that inform strategic decisions, enhancing the decision-making process within development teams~\cite{he_llm-based_2024}.

The integration of LLMs into cognitive agents empowers them to handle tasks that require understanding advanced linguistic patterns and domain knowledge, making them valuable assets in Agile software development environments~\cite{he_llm-based_2024}.

\begin{figure}[H]
  \includegraphics[width=\columnwidth,keepaspectratio]{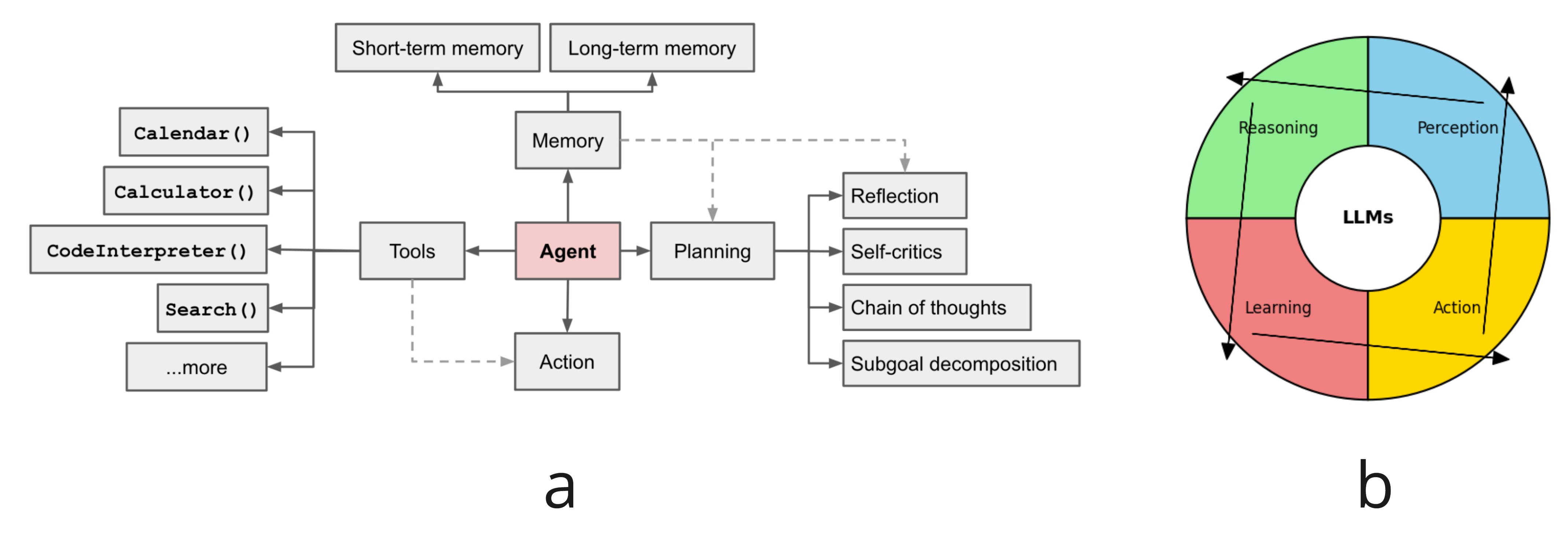}
  \caption{(\textbf{a}) Agent overview \cite{weng_llm_nodate}; (\textbf{b}) cognitive agent architecture represented as a cyclic process with four components.}
  \label{fig:Architectures_a_b}

\end{figure}

The integration of Large Language Models into cognitive agents enhances their linguistic capabilities while also advancing their overall cognitive functions through foundational components such as planning, memory, and tool use. In an LLM-powered autonomous system, as shown in Figure \ref{fig:Architectures_a_b}a, the LLM functions as the agent’s central processor, coordinating processes to efficiently address complex tasks. Planning involves breaking down extensive objectives into manageable subgoals, facilitating systematic advancement through structured strategies. Reflection and self-assessment enable the agent to analyze prior actions, learn from mistakes, and optimize future strategies, thereby increasing the precision and effectiveness of its outcomes.

Memory is structured into short-term and long-term functionalities. Short-term memory employs in-context learning and prompt engineering to quickly assimilate new information, while long-term memory supports the storage and retrieval of extensive information over extended durations, often using external vector stores and rapid-access systems. Furthermore, the agent's ability to utilize tools enhances its capabilities beyond the limitations of pretrained models. By integrating external APIs, real-time data, executable code, and proprietary information, the agent substantially improves its adaptability and operational efficiency in evolving environments.

Figure~\ref{fig:Architectures_a_b}b illustrates the architecture of cognitive agents, highlighting the four interdependent components, perception, reasoning, learning, and action, arranged in a cyclic process. At the center of this architecture, Large Language Models act as the unifying core, enhancing each layer's functionality. The perception layer utilizes LLMs to improve natural language understanding, enabling agents to process and interpret complex linguistic inputs. In the reasoning layer, LLMs provide contextual insights and support advanced decision-making processes. The learning layer benefits from the continuous learning capabilities offered by LLMs, enabling agents to adapt and evolve using textual data. Finally, the action layer leverages LLMs to facilitate language-based task execution, ensuring efficient and accurate interaction with the environment. The cyclic design emphasizes the iterative and interconnected nature of these components, driven by the central role of LLMs.
\vspace{-5pt}

{
\textls[-5]{Recent advancements in the application of Large Language Models demonstrate their potential in educational and project management contexts, as evidenced by the MCQGen framework. This framework leverages LLMs combined with retrieval-augmented generation and advanced prompt engineering to automate personalized content creation \cite{10577164}.}
}

\subsection{Multi-Agent Systems}

Multi-Agent Systems consist of interacting agents that collaborate to achieve individual or shared goals within an environment \mbox{\cite{wooldridge_introduction_2012}}. Each agent operates autonomously, perceiving its environment, making decisions, and taking actions. The agents in a MAS can be homogeneous—where all agents share identical capabilities, roles, and behavior—or heterogeneous, with agents differing in capabilities, roles, or objectives, allowing for specialization and complementary problem-solving. This distinction enables the MAS to be tailored to specific applications, with homogeneous systems offering simplicity and uniformity, while heterogeneous systems provide greater flexibility and complexity. In addition, their interactions can range from cooperative to competitive \cite{synergymas, paclic_elliottagents}.

\begin{figure}[H]
  \includegraphics[width=\columnwidth,keepaspectratio]{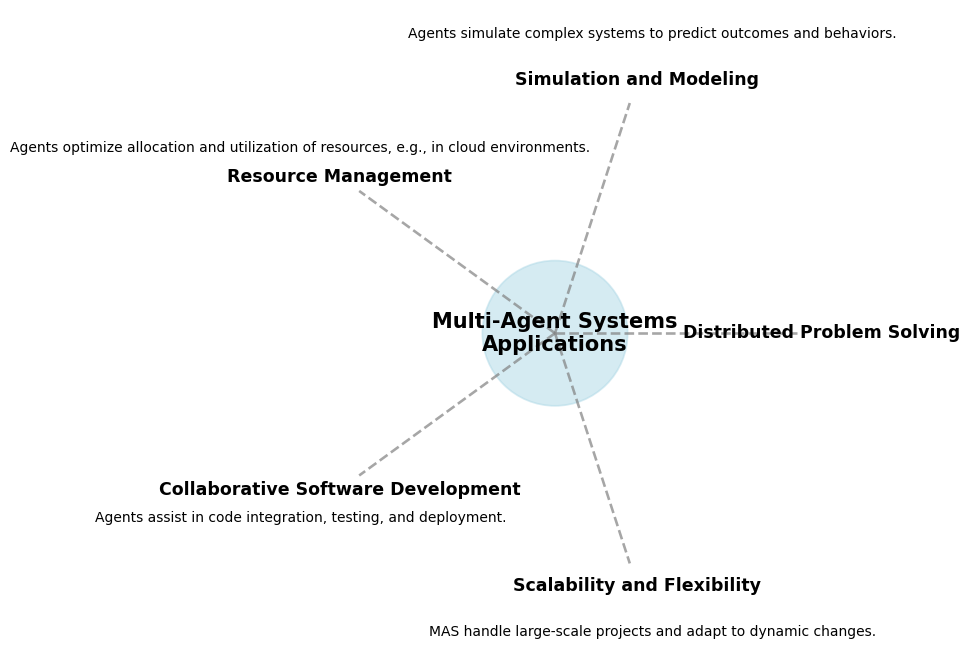}
  \caption{Applications of Multi-Agent Systems in software engineering.}
  \label{fig:mas_applications_wheel}
\end{figure}

In software engineering, Multi-Agent Systems are applied in various domains. \mbox{Figure~\ref{fig:mas_applications_wheel}} illustrates these applications, including distributed problem-solving, simulation and modeling, resource management, and collaborative software development.

In distributed problem-solving, agents divide complex problems into tasks and solve them concurrently, improving efficiency~\cite{durfee_trends_1989}. Simulation and modeling involve the use of MASs to replicate real-world systems, such as traffic networks or social behaviors, to analyze and predict outcomes~\cite{macal_tutorial_2010}. Resource management tasks are managed by agents that handle resources in cloud computing environments, optimizing allocation and utilization~\cite{Li2011}. In collaborative software development, agents help integrate, test, and deploy code, thereby enhancing collaboration between development teams~\cite{nguyen_agilecoder_2024}.
\vspace{-9pt}

The application of MASs in software engineering offers several benefits. Scalability is achieved because MASs can handle expanding systems by distributing tasks among agents, allowing the system to grow without a significant drop in performance~\cite{jennings_agent-based_2000}. Flexibility is provided by agents' ability to adapt to changes in the environment or requirements, making MASs suitable for systems where conditions frequently change~\cite{weiss_multiagent_2013}. Robustness is enhanced by the decentralized nature of the MAS, which reduces the single point of failure and increases the reliability of the system~\cite{ferber_multi-agent_1999}.

Frameworks such as ChatDev demonstrate the power of LLMs in unifying diverse roles through a structured, language-based communication model. By segmenting tasks into smaller, manageable phases and utilizing mechanisms like communicative dehallucination to ensure precision, these systems enhance the quality, completeness, and execution of generated software \cite{qian_chatdev_2024}.

Integrating cognitive agents powered by LLMs into MASs combines the benefits of both technologies, leading to systems capable of advanced reasoning, learning, and collaboration~\cite{cruz_transforming_2024}. This integration is particularly advantageous in Agile software development, where adaptability and enhanced communication are crucial.

\section{CogniSim Framework}
\label{sec:cognisim}

{
Before detailing its architecture, it is essential to clarify the nature of CogniSim and the capabilities it must encompass to address the challenges of large-scale Agile software development outlined in previous sections. CogniSim is a {{cognitive software framework}} designed to enhance and streamline Agile project management processes by integrating cognitive agents powered by Large Language Models into a Multi-Agent System environment. Rather than serving as a mere collection of standalone tools, CogniSim provides an intelligent system of LLM-augmented agents that collaborate with human teams.
To effectively address previously discussed challenges, such as increasing complexity, scalability, and the need for efficient communication, CogniSim must incorporate several key features and characteristics. These include cognitive assistance in understanding and reasoning about project contexts, the automation of labor intensive activities, inherent scalability aligned with Agile principles, and continuous quality assurance to ensure adherence to established practices. Table~\ref{tab:cognisim_features} summarizes these key attributes, highlighting how each directly addresses the challenges identified earlier.
}
\begin{table}[H]

\caption{{Key} features and characteristics of the CogniSim framework{.} 
}
\label{tab:cognisim_features}
\begin{adjustwidth}{-\extralength}{0cm}
\begin{tabularx}{\fulllength}{lX}
\toprule  
 \textbf{Feature/Characteristic} & \textbf{Description and Role in Addressing Challenges} \\ \midrule

Cognitive assistance & LLM-powered agents provide human-like reasoning capabilities, enabling them to interpret natural language requirements and adapt to changing project contexts. This aligns with previous analyses highlighting the need for intelligent automation in complex software projects~\cite{openai_gpt-4_2024, li_theory_2023}. \\ \midrule

Automation of routine tasks & By automating coding, documentation, and backlog refinement, the framework reduces human workload and cognitive overhead, allowing team members to focus on strategic decisions. Prior studies show that LLM-based code generation and documentation support improve productivity~\cite{chen_evaluating_2021, barua_exploring_2024}. \\ \midrule

Scalability and Agile alignment & CogniSim integrates seamlessly with Agile methodologies, particularly SAFe, ensuring synchronized development, continuous improvement, and effective communication across multiple teams and large-scale projects~\cite{safe6.0, shastri_role_2021}. \\ \midrule

Quality assurance and methodology adherence & Specialized agents continuously verify that deliverables meet coding standards, adhere to Agile processes, and align with strategic objectives. This ensures high-quality outputs, as evidenced by the importance of formal quality metrics in software engineering~\cite{fenton_software_2015, dingsoyr_decade_2012}. \\ \bottomrule

\end{tabularx}
\end{adjustwidth}
\end{table}
\vspace{-3pt}

{
With these foundational capabilities established, the following subsections detail the CogniSim framework’s layered architecture, its agent categorization, its integration with SAFe, and the quality measures that guide its performance evaluation.
}
\subsection{Framework Architecture}

\begin{figure}[H]

  \includegraphics[width=0.5\columnwidth,keepaspectratio]{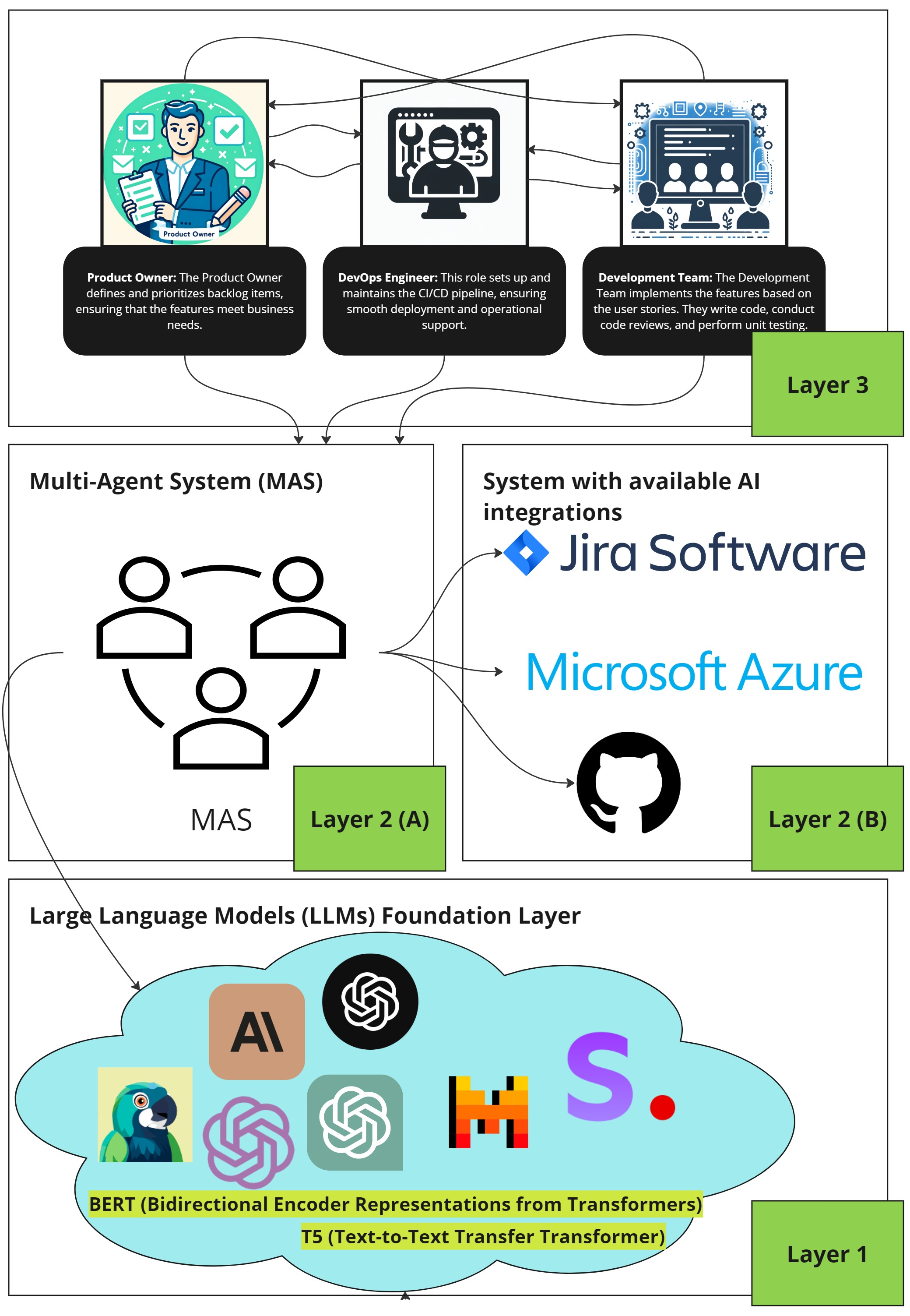}
  \caption{Layered ecosystem of CogniSim.}
  \label{fig:cognisim_layered_ecosystem}
\end{figure}

\textls[-15]{The CogniSim framework is built on a layered architecture that integrates cognitive agents, communication protocols, learning algorithms, decision-making frameworks, and collaboration tools. As depicted in Figure~\ref{fig:cognisim_layered_ecosystem}, the architecture consists of several key components. }

The Foundation Layer is composed of a Large Language Model, such as GPT, BERT, and T5, which serves as the backbone for natural language understanding, reasoning, and task execution. These models provide the foundational capabilities necessary for enabling higher-layer functionalities. This layer allows the system to process, analyze, and generate human-like responses, making it suitable for integration with various applications.

The Multi-Agent System Layer consists of two distinct components. Layer 2 (A) represents the core MAS framework where agents interact, collaborate, and make autonomous decisions. These agents leverage the LLM foundation for reasoning and adaptiveness. Layer 2 (B) encompasses systems with AI integrations, such as Jira, Microsoft Azure, and GitHub, which provide operational support and external APIs for seamless task management, development, and communication.

The Cognitive Agents Layer (Layer 3) includes roles such as Product Owners, DevOps Engineers, and Development Teams. These agents manage and streamline tasks using integrated AI systems. Product Owners define and prioritize backlog items, ensuring alignment with business needs. DevOps Engineers maintain the CI/CD pipeline for smooth deployments and operations. The Development Team implements features, conducts code reviews, and performs testing.

This modular and layered architecture ensures scalability, flexibility, and enhanced collaboration across human teams and AI-driven agents, optimizing software engineering~workflows.

\subsection{Agent Categorization}

Within the CogniSim framework, agents are categorized based on their roles and responsibilities to optimize collaboration and task execution. The main categories include Manager Agents, Executor Agents, Quality Checker Agents, and Methodology Reviewer Agents. Manager Agents are responsible for high-level decision-making, resource allocation, and overseeing project progress~\cite{Cinkusz2024}. These agents emulate roles such as Project Managers and Product Owners. Executor Agents focus on performing specific tasks such as coding, testing, and documentation~\cite{qiao_taskweaver_2024}, representing roles like Developers, QA Engineers, and Technical Writers. Quality Checker Agents ensure that deliverables meet predefined quality standards~\cite{shinn_reflexion_2023} by conducting code reviews, performing testing, and validating outputs against requirements. Methodology Reviewer Agents monitor adherence to Agile practices and methodologies~\cite{dingsoyr_decade_2012}, providing feedback on processes and suggesting improvements to enhance efficiency. Table~\ref{table:agent_roles} summarizes the agent categories and their primary functions within the framework.

Each agent is equipped with specific capabilities aligned with its role, enabling specialization and efficiency in task execution. The collaborative interaction among different agent types facilitates a comprehensive approach to project management, ensuring that all aspects of the software development lifecycle are effectively addressed.

\begin{table}[H]
  \caption{Agent categories and roles in CogniSim.}
  \label{table:agent_roles}
 
  \begin{tabularx}{\textwidth}{lL}
    \toprule
    \textbf{Agent Category} & \textbf{Primary Functions} \\ \midrule
    Manager Agents & Decision-making, resource allocation, project oversight \\ \midrule
    Executor Agents & Task execution (coding, testing, documentation) \\ \midrule
    Quality Checker Agents & Quality assurance, code reviews, output validation \\ \midrule
    Methodology Reviewer Agents & Process monitoring, adherence to Agile practices, feedback provision \\ \bottomrule
  \end{tabularx}
\end{table}

\subsection{Integration with SAFe}

The CogniSim framework aligns with the Scaled Agile Framework to enhance Agile practices in large-scale software development environments. The integration is achieved through several mechanisms. Firstly, cognitive agents are mapped to roles defined in SAFe, such as Release Train Engineers, Product Owners, and System Architects~\cite{safe6.0}. This role mapping ensures that agents fulfill responsibilities consistent with SAFe principles. Additionally, Manager Agents coordinate Executor and Quality Checker Agents to form virtual Agile Release Trains, facilitating synchronized development and delivery~\cite{knaster_safe_2020}. 

\begin{figure}[H]

  \includegraphics[width=0.85\columnwidth,keepaspectratio]{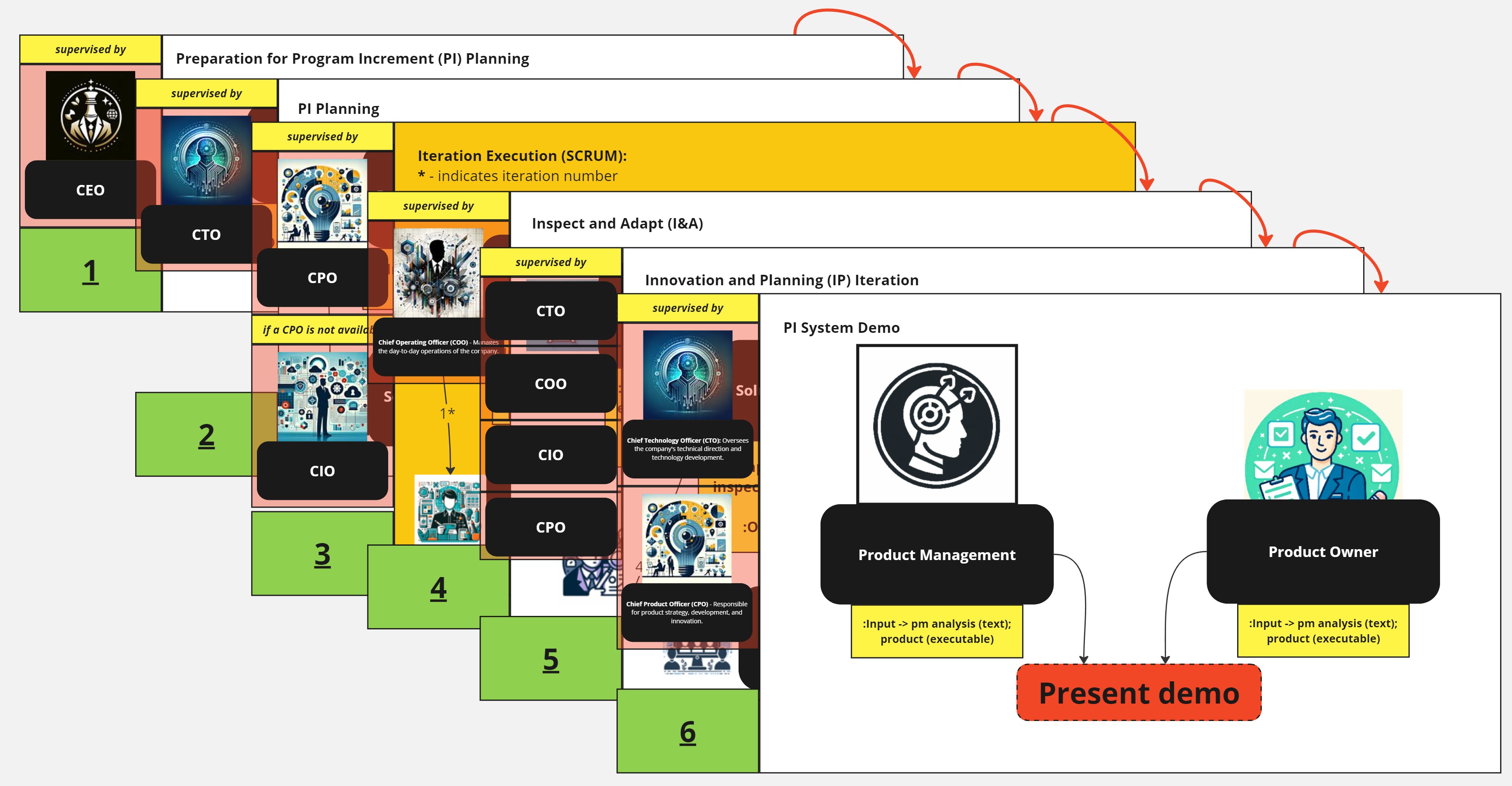}
  \caption{{Integration} of CogniSim with SAFe framework \cite{Cinkusz2024}{.} 
}
  \label{fig:cognisim_safe_integration}
\end{figure}

\textls[-15]{Methodology Reviewer Agents play a crucial role in promoting continuous improvement by monitoring processes and providing feedback, thereby fostering a culture of continuous improvement as advocated by SAFe~\cite{ebert_scaling_2017}. Furthermore, agents focus on delivering value by aligning their tasks with the organization's strategic objectives, enhancing customer satisfaction~\cite{shastri_role_2021}. Figure~\ref{fig:cognisim_safe_integration} illustrates how CogniSim integrates with the SAFe framework. }

By integrating with SAFe, CogniSim enhances coordination across multiple teams, improves communication efficiency, and ensures that development efforts are aligned with organizational goals. The cognitive agents automate routine coordination tasks, allowing human team members to focus on strategic decision-making and innovation.

{
Cognitive agents dynamically adjust their reasoning and decision-making processes as project priorities evolve, integrating newly identified requirements and stakeholder feedback into their ongoing workflows. By continuously reassessing user stories, backlog items, and architectural constraints, these agents reconfigure tasks, timelines, and resource allocations in response to changing conditions. This iterative adaptation ensures that cognitive agents maintain momentum, alignment, and flexibility within Agile environments, effectively supporting teams as project contexts change.
}
\section{Development Platform}
\label{sec4}

The CogniSim framework operates on a structured and adaptable development platform that integrates cognitive agents and Large Language Models into a Multi-Agent System. This section examines the platform's architectural design, its essential components, and the implementation features that underscore its flexibility and scalability for Agile software development practices.

\subsection{Platform Architecture}

\begin{figure}[H]

  \includegraphics[width=0.75\columnwidth,keepaspectratio]{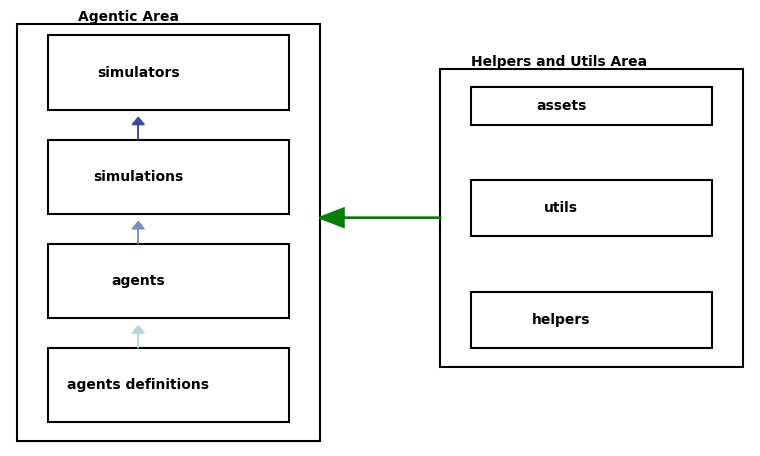}
  \caption{Project structure.}
  \label{fig:ProjectStructure}
\end{figure}

\textls[-15]{The CogniSim platform employs modular software engineering principles, ensuring efficient integration of cognitive agents and Large Language Models within a unified multi-agent framework. As depicted in Figure~\ref{fig:ProjectStructure}, the architectural approach prioritizes transparency, expandability, and system stability. The core directories, agents and agents\_definitions, organize the agents’ configurations, behaviors, and structural definitions. This logical segregation enhances the modularity of the system, facilitating effortless updates or enhancements to agent functionalities while preserving the integrity of the central system.}

The simulations and simulators directories form the backbone of the experimental environment, enabling the simulation of various scenarios and tasks. These components replicate real-world dynamics and support iterative testing, allowing agents to participate in software project management tasks by adhering to Agile methodologies. The helpers and utils directories provide utility functions that support the main simulation processes, such as data analysis, inter-agent communication, and task orchestration, thereby enhancing the platform’s flexibility and robustness.

The assets directory stores vital resources, including configuration files and datasets required for running the platform, ensuring that all necessary resources are readily available for simulation execution. Additionally, documentation files like README.md and requirements.txt offer comprehensive guidance on using the platform and managing dependencies, promoting clarity and reproducibility.

Overall, the architecture exemplifies an organized and scalable approach to developing cognitive agents capable of automating complex tasks in software project management. This modular design not only supports current simulations but also provides a solid foundation for future expansions, enabling more intricate simulations and a broader range of applications. By integrating advanced AI technologies and aligning with Agile methodologies, the CogniSim platform significantly enhances efficiency and productivity in software project management.

\subsection{Key Components and Functionalities}

The CogniSim platform is built using {Python version 3.11.10,} 
 selected for its extensive ecosystem of libraries that are particularly well suited for artificial intelligence, LLMs, and data manipulation tasks~\cite{langchain_introduction}. The Integrated Development Environment utilized is Visual Studio Code (VS Code version 1.96.2), chosen for its versatility and strong community support, which facilitates the handling of complex simulations and Multi-Agent Systems~efficiently.

\textls[-15]{A pivotal component of the platform is LangChain version 0.2, a framework designed to streamline the integration of LLMs into the Multi-Agent System. LangChain provides a modular and flexible environment that allows developers to combine various components, such as language models, prompts, memory modules, and external data sources, into cohesive workflows~\cite{langchain_core}. This framework is essential for managing the interactions between agents and leveraging the full potential of LLMs in generating context-aware and intelligent responses.}

\textls[-15]{The platform also incorporates OpenAI’s GPT-4 and GPT-3.5 models, selected for their advanced natural language processing capabilities, which are critical for enabling cognitive agents to perform high-level decision-making and communication~\cite{openai_gpt-4_2024}. Additionally, the platform leverages a variety of Python libraries, including tqdm for progress monitoring, pandas for data manipulation, websockets for real-time communication, and jsonschema for validating JSON structures, among others. These libraries collectively facilitate diverse functionalities such as monitoring simulation progress, managing structured data, enabling bidirectional communication between agents, and ensuring data integrity within the system.}

\subsection{Implementation Details}

The CogniSim platform is designed with a modular and flexible architecture to support dynamic multi-agent simulations. At its core, the platform emphasizes abstraction and scalability, enabling users to configure and run simulations without requiring direct interaction with underlying code.

Agents in the platform are defined using a high-level configuration format, allowing their roles, behaviors, and interactions to be easily specified. This approach provides a clear separation between agent definitions and the core system, ensuring adaptability across diverse simulation scenarios. For example, roles such as Chief Technology Officer or Solution Architect can be incorporated seamlessly into simulations, demonstrating the platform's capacity for managing complex, role-specific interactions.

Simulations are orchestrated through a structured process that involves defining the simulation environment, setting parameters such as iteration limits and objectives, and ensuring all interactions are systematically recorded for analysis. The platform employs tools to track and document every simulation run, producing detailed logs and summaries for further evaluation. These outputs can be presented in user-friendly formats to aid in understanding the interactions and decision-making processes within the simulation.

\section{Simulation Breakdown}
\label{sec5}

This section provides an examination of the CogniSim framework's mechanics, focusing on the setup, agent interactions, and output generation processes. This substantial analysis ensures consistency and clarity in simulation execution while offering substantial data for evaluating the framework's effectiveness in Agile software project management.

\subsection{Simulation Setup}

The simulation environment within the CogniSim framework is configured to emulate concrete software development processes adhering to Agile methodologies. The setup begins with defining key simulation parameters, including the simulation name, maximum iterations, and elements that guide the simulation's execution. These parameters establish the scope and boundaries of each simulation run, ensuring consistency and clarity in the simulation outcomes. Agents are initialized based on their roles, which are defined through JSON configuration files (Box~\ref{box:product-management-agent}) that 
outline their behaviors and interaction patterns. This initialization process is crucial for creating a tangible simulation environment where each agent operates according to its assigned responsibilities.

\begin{boxenv}[H]
\caption{{JSON} structure defining the Product Management Agent{.} 
}
\label{box:product-management-agent}

\begin{adjustwidth}{-\extralength}{0cm}
\fcolorbox{orange}[HTML]{fff2e6}{\small\parbox{\fulllength-2\fboxsep}{
  \parindent0mm 
\section*{{Agent Information} 
}
\begin{itemize}
    \item \textbf{Agent Name:} Product Management Agent---Alex
\end{itemize}

\section*{Agent Definition}
\subsection*{Prompt}
\begin{itemize}
    \item \textbf{Input Variables:}
    \begin{itemize}
        \item client\_analysis
        \item solution\_architect\_feedback
        \item instruction
    \end{itemize}
    \item \textbf{Template Messages:}
    \begin{itemize}
        \item \textbf{System:} As a Product Management professional, leverage your expertise in strategic planning and your deep understanding of market trends, customer needs, and the competitive landscape to develop a comprehensive vision and road map for the product. Your plan should align with the overall business strategy and customer expectations while being informed by the technical insights provided by the solution architect. Please adhere to the following steps:
        \begin{enumerate}
            \item Analyze the detailed client analysis and solution architect feedback to inform your strategic direction.
            \item Define the product vision that aligns with the business strategy and customer needs.
            \item Develop a detailed road map and objectives for the upcoming Program Increment (PI), outlining the features and capabilities to be developed.
            \item Articulate the business context, including market changes and strategic goals, to justify the chosen direction.
            \item Engage with Product Owners and the solution architect to ensure alignment and address any technical or market-related feedback.
            \item Communicate the strategic plan to all stakeholders, ensuring clarity and buy-in.
            \item Establish metrics for success and outline how progress will be measured against the objectives.
        \end{enumerate}
        \item \textbf{User:} This is the client analysis \{client\_analysis\}.
        \item \textbf{User:} This is the solution architect feedback \{solution\_architect\_feedback\}.
        \item \textbf{Placeholder:} instruction
    \end{itemize}
    \item \textbf{Template Format:} f-string
\end{itemize}

\subsection*{LLM}
\begin{itemize}
    \item \textbf{Model Name:} gpt-3.5-turbo
    \item \textbf{Temperature:} 0
    \item \textbf{Max Tokens:} null
\end{itemize}

\section*{Tools}
\begin{itemize}
    \item \textbf{Allowed Basic Tools:} None
    \item \textbf{Allowed External Tools:}
    \begin{itemize}
        \item TavilySearchResults
    \end{itemize}
\end{itemize}

\section*{Return Values}
\begin{itemize}
    \item \textbf{Output:}
    \begin{itemize}
        \item output
    \end{itemize}
\end{itemize}

\section*{Agent Type}
\begin{itemize}
    \item \textbf{Type:} dialogue\_agent\_with\_tools
\end{itemize}}}
\end{adjustwidth}
\end{boxenv}

To provide further clarity, the JSON structure used for defining the Product Management Agent is included below. 
This example captures the essential parameters, including the prompt, input variables, roles, and allowed tools, which govern the agent's behavior and interactions within the simulation.

Environment \textls[+15]{variables are loaded to configure the simulation context, allowing agents to access necessary resources and data required for task execution. The JSON structure defines critical inputs, such as client analysis and solution architect feedback, which serve as starting points for simulations and provide agents with a clear understanding of project objectives and constraints. In addition, description files offer step-by-step task guides, delineating the sequence of interactions and ensuring alignment with Agile methodologies. This structured setup enables a controlled yet adaptable simulation environment, allowing agents to engage in complex interactions that mimic real-world project management~scenarios}.

\subsection{Agent Interactions}

In the simulation, agents interact through systematically organized communication protocols that facilitate structured dialogue and collaboration. The orderly progression of exchanges is maintained by a speaker selection mechanism, where agents employ their Large Language Models to produce contextually relevant responses that influence decision-making processes effectively.

Each agent maintains a record of past interactions, ongoing tasks, and project statuses, ensuring both continuity and adaptability. This functionality enables the CogniSim framework to replicate intricate interactions, offering a reliable environment for evaluating cognitive agents within the context of Agile project management.

\subsection{Chat History Output}

The Chat History Output Visualization (Boxs~\ref{box:iteration-execution} and~\ref{box:preparing-for-program-increment}) provides a detailed view of the communication dynamics between cognitive agents during the simulation. This visualization highlights the flow of messages, enabling a closer analysis of agent interactions, decision-making processes, and collaborative behavior. Messages are presented with attribution to individual agents, truncated to maintain readability, and formatted to ensure clarity. By reviewing this output, readers can better understand the practical implementation of dialogue-driven simulations and how agents engage in real-time problem-solving within the Agile framework.

\begin{boxenv}[H]

\caption{{Chat}
 History---Iteration Execution.}
 \label{box:iteration-execution}
\begin{adjustwidth}{-\extralength}{0cm}
\fcolorbox[HTML]{006060}[HTML]{e6f2f2}{\small\parbox{\fulllength+5\fboxsep}{
  \parindent0mm 

\begin{description}
  \item[UX Designers:]
  \begin{quote}
  During the Iteration Execution phase, UX Designers collaborate with the Development Team to refine designs, adjust specifications, and iterate on assets based on feedback and technical constraints.
  \end{quote}

  \item[DevOps Engineer:]
  \begin{quote}
  As the DevOps Engineer, ensure CI/CD pipelines support integration, testing, delivery, and automation to enhance the efficiency of builds, deployments, and troubleshooting processes with Development Teams.
  \end{quote}

  \item[System Team:]
  \begin{quote}
  The System Team supports integration and testing environments, resolves infrastructure issues, and collaborates with DevOps Engineers to maintain stability and address changes during Iteration Execution.
  \end{quote}

  \item[Development Team:]
  \begin{quote}
  The Development Team focuses on user stories, communicates design needs with UX Designers, ensures environment readiness with System Team, and raises dependencies or issues during the iteration.
  \end{quote}

\end{description}}}

\end{adjustwidth}
\end{boxenv}


\begin{boxenv}[H]

\caption{{Chat} 
 History---preparing for program increment.}
 \label{box:preparing-for-program-increment}
\begin{adjustwidth}{-\extralength}{0cm}
\fcolorbox[HTML]{006060}[HTML]{e6f2f2}{\small\parbox{\fulllength+5\fboxsep}{
  \parindent0mm 

\begin{description}
  \item[Product Management---Alex:]
  \begin{quote}
  System Architect, can you outline a high-level approach for using a Bayesian network to improve our real-time trading decisions on Coinbase?
  \end{quote}

  \item[System Architect---John:]
  \begin{quote}
  Great, let us dive a bit deeper into each step. Starting with the first one, Data Gathering, what kind of historical data are we talking about? How far back should we go? Furthermore, how do we ensure the real-time data are accurate and timely for making predictions?
  \end{quote}

  \item[System Architect---John:]
  \begin{quote}
  That makes sense. Now, looking at the second step, could you explain further how we define the structure and variables of the Bayesian network? What factors should we consider when deciding on the structure? Furthermore, how do we determine the relevant variables?
  \end{quote}

  \item[System Architect---John:]
  \begin{quote}
  I see, so the network structure is dynamic and might need adjustments over time. Moving on to the third step, Learning Parameters, could you elaborate a bit more on how we learn these parameters? What techniques or algorithms could we use? Furthermore, how do we validate the accuracy of these parameters?
  \end{quote}

  \item[System Architect---John:]
  \begin{quote}
  Thank you for explaining that. Now let us discuss the fourth step, Inference. Could you explain how this process works in detail? How do we use the learned parameters to make predictions? Furthermore, how do we handle uncertainty in the prediction process?
  \end{quote}

  \item[Product Management---Alex:]
  \begin{quote}
  Thank you for all the details provided so far. Now that we have refined the approach, could you please provide a final consolidated list of the key tasks, roles, and next steps for implementing this Bayesian network integration?
  \end{quote}

  \item[System Architect---John:]
  \begin{quote}
  Absolutely. The final plan for implementing the Bayesian network for real-time crypto trading on Coinbase is as follows:
  (\ldots)
  \end{quote}
\end{description}}}
\end{adjustwidth}
\end{boxenv}

\subsection{Output Analysis}

The outputs generated by the simulations provide substantial data for evaluation and analysis. Each simulation run produces chat logs that capture the full interaction between agents, documenting the flow of communication, decision-making processes, and task executions. These logs are saved in both HTML and text formats, offering different levels of accessibility and readability for various analytical purposes.

HTML files present the chat history in a structured and visually clear format, with color-coded messages and agent-specific icons enhancing the clarity of interactions. This visualization aids in quickly identifying patterns, key decision points, and the effectiveness of communication among agents. Text files provide a straightforward and unformatted record of the dialogues, suitable for in-depth textual analysis and comparison against predefined benchmarks.

Execution logs capture the simulation's runtime details, including timestamps, system interactions, and any warnings or errors encountered during execution. These logs are essential for debugging purposes and for understanding the underlying processes that drive agent interactions. Additionally, configuration files documenting the input parameters and agent setups are generated, ensuring that each simulation run is reproducible and that the conditions under which the simulation was conducted are well documented.

The analysis of these outputs involves assessing the performance of agents based on predefined metrics such as task completion time, quality of deliverables, and communication efficiency. By evaluating these metrics, insights can be gained into the strengths and limitations of the CogniSim framework, as well as the effectiveness of LLM-powered cognitive agents in managing Agile software development tasks. This substantial output analysis is critical for validating the framework's capabilities and for identifying areas for further improvement and optimization.

\subsubsection{Visualization of Results}

The HTML output files generated by the simulations provide a visual representation of agent interactions, which is essential in assessing the flow and coherence of communication. Each agent's messages are color-coded and accompanied by icons, allowing for quick differentiation and analysis of individual contributions within the dialogue. This visual format facilitates the identification of communication patterns, highlighting how agents coordinate tasks, resolve conflicts, and make collective decisions. By reviewing these visual logs, researchers can gain a deep understanding of the agents' collaborative dynamics and the overall effectiveness of the simulation in replicating concrete Agile interactions.

\subsubsection{Log Analysis}

Execution logs play a pivotal role in the simulation breakdown by offering a complete trace of the simulation's execution flow. These logs include timestamps, system actions, and any warnings or errors that occur, providing significant insights into the functional aspects of the simulation. For example, deprecation warnings and other system-level messages can indicate potential issues with the codebase or the need for updates to dependencies. Analyzing these logs helps in identifying and troubleshooting technical problems, ensuring the robustness and reliability of the simulation environment. Moreover, the logs capture the sequence of agent actions and responses, enabling a thorough examination of the agents' decision-making processes and their adherence to the simulation's objectives.

\subsubsection{Reproducibility}

Ensuring reproducibility is a fundamental aspect of the CogniSim framework, achieved by completely documenting all input parameters and agent configurations used in each simulation run. Configuration files in JSON format store full information about agent roles, behaviors, and interaction patterns, allowing simulations to be replicated accurately under identical conditions. This comprehensive documentation enables researchers to validate results, conduct repeated experiments, and compare outcomes across different simulation runs. Reproducibility is essential for establishing the reliability of the framework and for facilitating continuous improvement based on consistent and comparable data.

Overall, the simulation breakdown provides a substantial overview of how the CogniSim framework operates, detailing the setup, agent interactions, and output generation processes. This analysis ensures that each simulation run is executed in a controlled and consistent manner, providing significant data for evaluating the framework's effectiveness in enhancing Agile software project management through cognitive agents powered by Large Language Models. 
\section{Case Study}
\label{sec:case_study}

{
This case study aims to address this specific research question: {{``How effectively can cognitive agents, powered by LLMs, replicate and enhance key roles and processes within Agile (SAFe) software development environments?''}} While this new question complements and builds upon the overarching research questions (RQ1 and RQ2) introduced earlier, our immediate focus here is on evaluating the CogniSim framework against defined quality characteristics—namely, decision-making effectiveness, communication efficiency, and adaptability under realistic project constraints. The findings from this chapter will serve as groundwork for a more detailed reflection on RQ1 and RQ2, which will be revisited and discussed at length in the next chapter.
}

\subsection{Case Study Overview}

The simulation focuses on a comprehensive software project that involves the integration of advanced APIs and the development of comprehensive business logic. The roles simulated within this environment are described in Section \ref{sec:cognisim}. By modeling these interactions, the case study aims to showcase how the CogniSim framework can effectively handle real-world software engineering challenges, such as aligning technical capabilities with business objectives, managing dependencies, and maintaining project agility.

\subsection{Case Study Approach}

The methodology for this case study employs a qualitative research approach, utilizing the CogniSim framework to simulate interactions among various astute agents. The process begins with the simulation setup, where the project scope, objectives, and agent roles are meticulously defined based on a realistic software development scenario. Agents are then configured using JSON files that specify their roles, behaviors, and responsibilities, ensuring a high level of customization and scalability.
{
Data collection occurs over multiple simulation runs corresponding to distinct SAFe phases. In each run, we log agent communications, decisions, and outcomes (e.g., features implemented, technical decisions made). We capture the following:
\begin{itemize}
    \item {{Agent dialogue and decisions:}} All agent-to-agent and agent-to-environment messages are recorded, providing a complete trace of negotiation, planning, and execution activities.
    \item {{Performance metrics:}} We measure task completion times and adherence to project timelines.
\end{itemize}
}
Following the configuration, the simulation is executed through multiple iterations to emulate the cyclical nature of Agile processes, including Program Increment (PI) Planning, Iteration Execution, and Retrospectives. During each iteration, data are collected on agent interactions, task completions, and decision-making processes. These data are then analyzed to evaluate the performance of the agents against predefined metrics such as task completion time, quality of deliverables, and communication efficiency.

\begin{figure}[H]

  \includegraphics[width=\columnwidth,keepaspectratio]{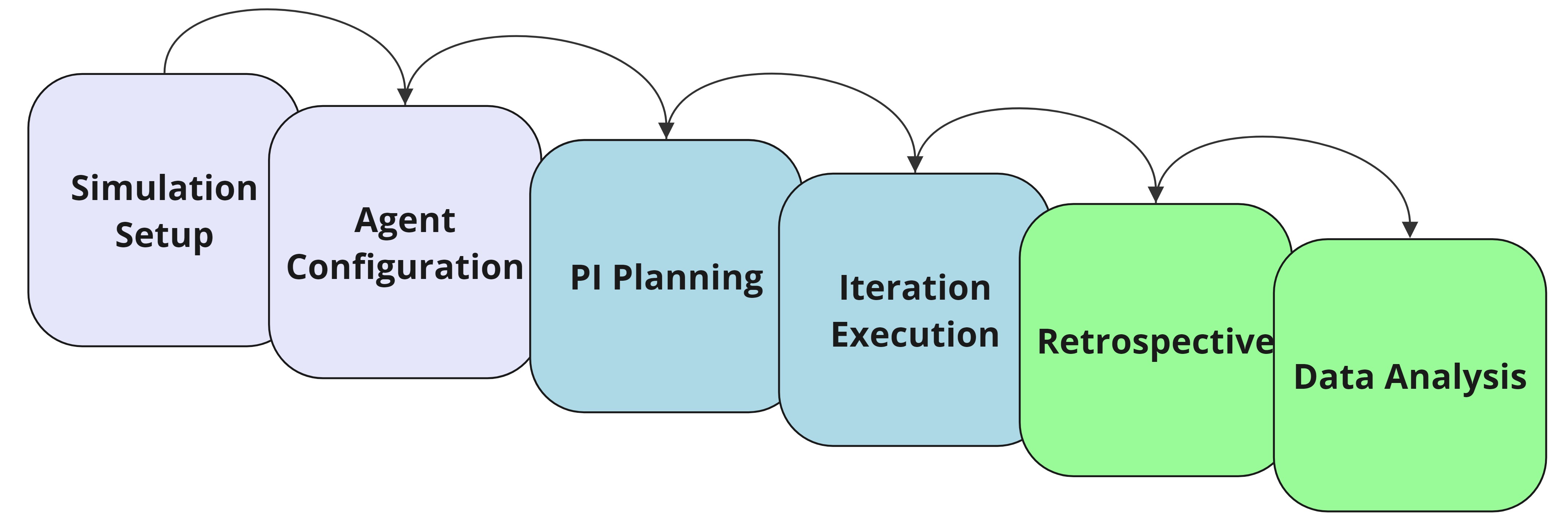}
  \caption{Simulation workflow in CogniSim, showing the iterative Agile process from setup through data analysis.}
  \label{fig:simulation_workflow}
\end{figure}

The simulation workflow is illustrated in Figure~\ref{fig:simulation_workflow}, which shows the iterative Agile process from setup through data analysis.

\subsection{Results and Analysis}

The simulation results provide compelling evidence of the CogniSim framework’s effectiveness in enhancing Agile software development processes. One of the most significant findings is the enhancement in decision-making capabilities of the agents. Astute agents demonstrated the ability to analyze advanced scenarios and make informed decisions swiftly, leading to more efficient problem-solving and task management. This was particularly evident in the PI Planning phase, where agents successfully negotiated feature priorities and identified technical dependencies, ensuring that project objectives were aligned with both business goals and technical feasibility.

Efficiency improvements were another significant outcome, with task completion times being significantly reduced compared to human-managed processes. The agents’ ability to process information rapidly and execute tasks without the typical bottlenecks associated with human teams contributed to this increased efficiency. Additionally, the consistency and reliability of the agents' outputs were enhanced, as they maintained a high level of quality and minimized errors throughout the simulation.

{
Agents continued to communicate effectively, and the distribution of tasks across multiple agents reduced the load on individual components without increasing coordination overhead. Although these results are preliminary and gathered from controlled simulations, they indicate that the CogniSim framework can grow in complexity and team size without a corresponding drop in performance.
}

Overall, the results indicate that the CogniSim framework can effectively augment Agile software development practices by automating routine tasks, enhancing decision-making processes, and fostering efficient collaboration among team members. These improvements contribute to a more streamlined and productive software development lifecycle, demonstrating the potential of integrating astute agents powered by LLMs into project management scenarios.

\subsection{Key Insights and Applications}

The case study emphasizes several important insights regarding the integration of perceptive agents in Agile software project management. Firstly, the scalability of the CogniSim framework is notable, as it supports larger teams and advanced projects without a corresponding increase in resource requirements. {Specifically, by incrementally increasing both the number of agents and project complexity, we observed stable task completion rates, manageable communication overhead, and consistent decision-making quality, thereby providing preliminary evidence for scalability.}

Secondly, the adaptability of the agents proves essential in dynamic project environments. Their capacity to address shifting requirements and evolving project conditions ensures that the development process remains agile and attuned to stakeholder needs. This adaptability is particularly advantageous in circumstances requiring swift responses to unexpected challenges or the incorporation of new features.

Lastly, the collaboration between human team members and perceptive agents underscores the synergistic strengths of both; while agents efficiently automate routine tasks and manage large volumes of data, human oversight is indispensable for strategic decision-making and addressing complex issues that demand contextual understanding and innovative problem-solving. This effective partnership between humans and perceptive agents enhances project management by leveraging their combined strengths to achieve superior outcomes in software development.

{
In summary, through a structured qualitative methodology, defined research questions, thematic data analysis, illustrative dialogue snippets, and performance measurements, this case study provides a more rigorous demonstration of the CogniSim framework’s capabilities. Future work will involve more extensive quantitative analyses and real-world field studies to further validate these findings.}
\section{Experiments and Results}
\label{sec:experiments_results}

{
Building upon the research questions outlined in Section~\ref{sec:introduction}, this section evaluates the CogniSim framework’s effectiveness in achieving the previously stated objectives. Rather than restating those questions here, we focus on how well the framework addresses the identified challenges and objectives presented in the introduction.
}

{
To address these questions, we focus on evaluating specific characteristics such as performance, quality of deliverables, and collaborative effectiveness. Performance is measured through metrics like task completion time and backlog reduction, reflecting the efficiency of the system. Quality of deliverables considers factors such as adherence to coding standards and the clarity of generated documentation, emphasizing the reliability and usability of outputs. Collaborative effectiveness assesses aspects like communication efficiency and adaptability, highlighting the system's ability to facilitate teamwork and respond to dynamic scenarios. These attributes act as dependent variables, while independent variables include the choice of LLM model (e.g., GPT-3.5-turbo versus GPT-4), the number of simulation iterations, agent role configurations, and prompt parameters such as temperature and memory settings (as shown at Box~\ref{box:simulation-configruation}). By examining these variables, we aim to analyze the interplay between system inputs and outcomes.
}

\begin{boxenv}[H]

\caption{{JSON} 
 structure for simulation configuration.}
 \label{box:simulation-configruation}
\begin{adjustwidth}{-\extralength}{0cm}
\fcolorbox[HTML]{0000bf}[HTML]{e6e6ff}{\small\parbox{\fulllength-2\fboxsep}{
\parindent0mm

\section*{Simulation Configuration}
\begin{itemize}
    \item \textbf{Simulations:}
    \begin{itemize}
        \item Simulation 1:
        \begin{itemize}
            \item \textbf{sim\_id:} 1
            \item \textbf{model\_type:} gpt-3.5-turbo
            \item \textbf{iterations:} 10
            \item \textbf{temperature:} 0.7
            \item \textbf{agents\_involved:}
            \begin{itemize}
                \item Product Management
                \item System Architect
            \end{itemize}
        \end{itemize}
        \item Simulation 2:
        \begin{itemize}
            \item \textbf{sim\_id:} 2
            \item \textbf{model\_type:} gpt-4
            \item \textbf{iterations:} 10
            \item \textbf{temperature:} 0.7
            \item \textbf{agents\_involved:}
            \begin{itemize}
                \item Product Management
                \item System Architect
            \end{itemize}
        \end{itemize}
    \end{itemize}
\end{itemize}}}
\end{adjustwidth}
\end{boxenv}
\vspace{-3pt}

{
In this experimental setup, the cognitive agents collectively simulate the development of a hypothetical enterprise-level software solution---such as a web-based application integrating external payment APIs, user management modules, and advanced data analytics features. As a result, agents produce various software artifacts: code snippets that implement API integrations, user stories refined into actionable backlog items, architectural decision logs, UI/UX design recommendations, and deployment pipeline configurations. By analyzing these outputs, we gain insight into the quality and completeness of the software engineering process facilitated by CogniSim.
}

\subsection{Experimental Design}

The experiments conducted within the CogniSim framework were designed to evaluate the performance and effectiveness of cognitive agents in simulating Agile software development processes. {These simulations replicate different SAFe phases—Preparation for Program Increment (PI) Planning, PI Planning, Iteration Execution, Inspect-and-Adapt Workshop, Innovation and Planning (IP) Iteration, and PI System Demo—thereby providing multiple data points to assess how agents behave under varying conditions and project~stages.}

{
The \textls[-20]{independent variables manipulated during these experiments include the following:}
\begin{itemize}
    \item {{Model type (independent variable):}} GPT-3.5-turbo or GPT-4.
    \item {{Number of iterations (independent variable):}} Variable between runs (e.g., 10, 50) to observe long-term behavior.
    \item {{Agent roles (independent variable):}} Adjusting which roles are included (Product Management, System Architect, Development Team, etc.).
    \item {{Temperature and prompt settings (independent variable):}} Influencing the creativity and precision of agent outputs.
\end{itemize}
}
{
The dependent variables include the following:
\begin{itemize}
    \item {{Performance metrics:}} Task completion time, backlog reduction rate.
    \item {{Quality metrics:}} Code adherence to standards, clarity of documentation, correctness of architectural decisions.
    \item {{Collaboration and communication metrics:}} Frequency and quality of agent interactions, consistency in decision-making, adaptability to changing requirements.
\end{itemize}}

The experimental design encompassed six cohesive simulations, each representing a pivotal SAFe phase, assigning roles to agents such as Product Managers, System Architects, Development Teams, UX Designers, and DevOps Engineers. By running these simulations through multiple iterations, the experiments aimed to capture the cyclical nature of Agile processes, allowing for a cohesive assessment of agent interactions, task management, and decision-making efficacy.

\subsubsection*{Agent Parameters for Experimental Variations}

To comprehensively assess the impact of pivotal factors on the performance of cognitive agents, a range of agent parameters were systematically varied across simulation runs. Table~\ref{tab:agent-parameters} outlines the key agent parameters adjusted during the experiments.

\begin{table}[H]

\caption{{Agent} parameters.}
\label{tab:agent-parameters}
\begin{adjustwidth}{-\extralength}{0cm}
\begin{tabularx}{\fulllength}{lLL}
\toprule
\textbf{Parameter} & \textbf{Description} & \textbf{Possible Values} \\
\midrule
Model type & The language model used for agent responses. & GPT-3.5-turbo, GPT-4. \\
\midrule
Number of iterations & Number of turns or exchanges between agents. & Any positive integer (e.g., 10, 50, 100). \\
\midrule
Temperature & Controls randomness in the model's output (creativity level). & 0.0 (deterministic) to 1.0 (maximum randomness) \cite{peeperkorn_is_2024}. \\
\midrule
Max tokens & Maximum length of responses from agents. & Any positive integer (e.g., 150, 500). \\
\midrule
Agent roles & Different roles or agents involved in the simulation. & Product Management, System Architect, Development Team, QA Engineer, etc. \\
\midrule
Prompt templates & Different initial prompts or instructions for agents. & Varied prompts per agent to test impact on responses. \\
\midrule
Input variables & Specific input data provided to agents (e.g., client analysis, objectives). & Different scenarios or datasets for testing. \\
\midrule
Selection function & Method for selecting the next speaker in the simulation. & Alternate speakers, random selection, directed selection. \\
\midrule
Elaboration functions & Use of functions that elaborate or expand on topics (e.g., topic elaboration). & Enabled, disabled. \\
\midrule
API parameters & Other OpenAI API parameters like presence\_penalty, frequency\_penalty. & presence\_penalty: $-$2.0 to 2.0; frequency\_penalty: $-$2.0 to 2.0. \\
\midrule
Agent memory & Amount of prior conversation history agents remember. & Full memory, limited memory (e.g., last 3~messages). \\
\midrule
Agent personality or style & Communication style of agents (affects language used). & Formal, casual, technical, persuasive. \\
\bottomrule
\end{tabularx}
\end{adjustwidth}
\end{table}
\newpage
An example setup for simulations with varied simulation parameters is presented in Table~\ref{tab:simulation-examples}.

\begin{table}[H]

\caption{{Example} simulations with varied parameters.}
\label{tab:simulation-examples}
\begin{adjustwidth}{-\extralength}{0cm}
\begin{tabularx}{\fulllength}{cLccLL}
\toprule
\textbf{Sim ID} & \textbf{Model Type} & \textbf{Iterations} & \textbf{Temperature} & \textbf{Agents Involved} & \textbf{Notes} \\
\midrule
1 & GPT-3.5-turbo & 10 & 0.7 & Product Management, System Architect & Baseline simulation \\
\midrule
2 & GPT-4 & 10 & 0.7 & Product Management, System Architect & Testing with GPT-4 \\
\midrule
3 & GPT-3.5-turbo & 50 & 0.7 & Product Management, System Architect & Increased iterations \\
\midrule
4 & GPT-3.5-turbo & 10 & 0.5 & Product Management, System Architect & Lower temperature (less randomness) \\
\midrule
5 & GPT-3.5-turbo & 10 & 0.9 & Product Management, System Architect & Higher temperature (more randomness) \\
\midrule
6 & GPT-3.5-turbo & 10 & 0.7 & Product Management, System Architect, Dev Team & Added Development Team agent \\
\bottomrule
\end{tabularx}
\end{adjustwidth}
\end{table}
\vspace{-3pt}

By varying these parameters, this study aimed to investigate their effects on the agents' performance, interaction dynamics, and the overall quality of the simulations. The systematic variation of parameters allowed exploration of different scenarios and configurations, providing insights into optimal settings for cognitive agent simulations within the Agile framework.

\subsection{Results and Analysis}

The results of the experiments are captured in Table~\ref{tab:simulation_results}, detailing the performance metrics obtained during the simulations. The table includes key metrics such as unique content percentage, diversity score, completion score, and sentiment stability for each simulation run.

\begin{table}[H]

\caption{{Simulation} results.}
\label{tab:simulation_results}
\begin{adjustwidth}{-\extralength}{0cm}
\begin{tabularx}{\fulllength}{cl>{\centering\arraybackslash}X>{\centering\arraybackslash}X>{\centering\arraybackslash}X>{\centering\arraybackslash}X}
\toprule
\textbf{Sim ID} & \textbf{Model Type} & \textbf{Unique Content} & \textbf{Diversity Score} & \textbf{Completion Score} & \textbf{Sentiment Stability} \\
\midrule
1 & gpt-3.5-turbo & 100.00 & 0.45 & 50.00 & 0.00 \\
\midrule
2 & gpt-4 & 100.00 & 0.77 & 0.00 & 100.00 \\
\midrule
3 & gpt-3.5-turbo & 83.67 & 0.43 & 50.00 & 100.00 \\
\midrule
4 & gpt-3.5-turbo & 11.11 & 0.46 & 0.00 & 100.00 \\
\midrule
5 & gpt-3.5-turbo & 33.33 & 0.46 & 0.00 & 100.00 \\
\midrule
6 & gpt-3.5-turbo & 100.00 & 0.63 & 50.00 & 33.33 \\
\bottomrule
\end{tabularx}
\end{adjustwidth}
\end{table}
\vspace{-3pt}

\begin{figure}[H]

\includegraphics[width=0.85\textwidth]{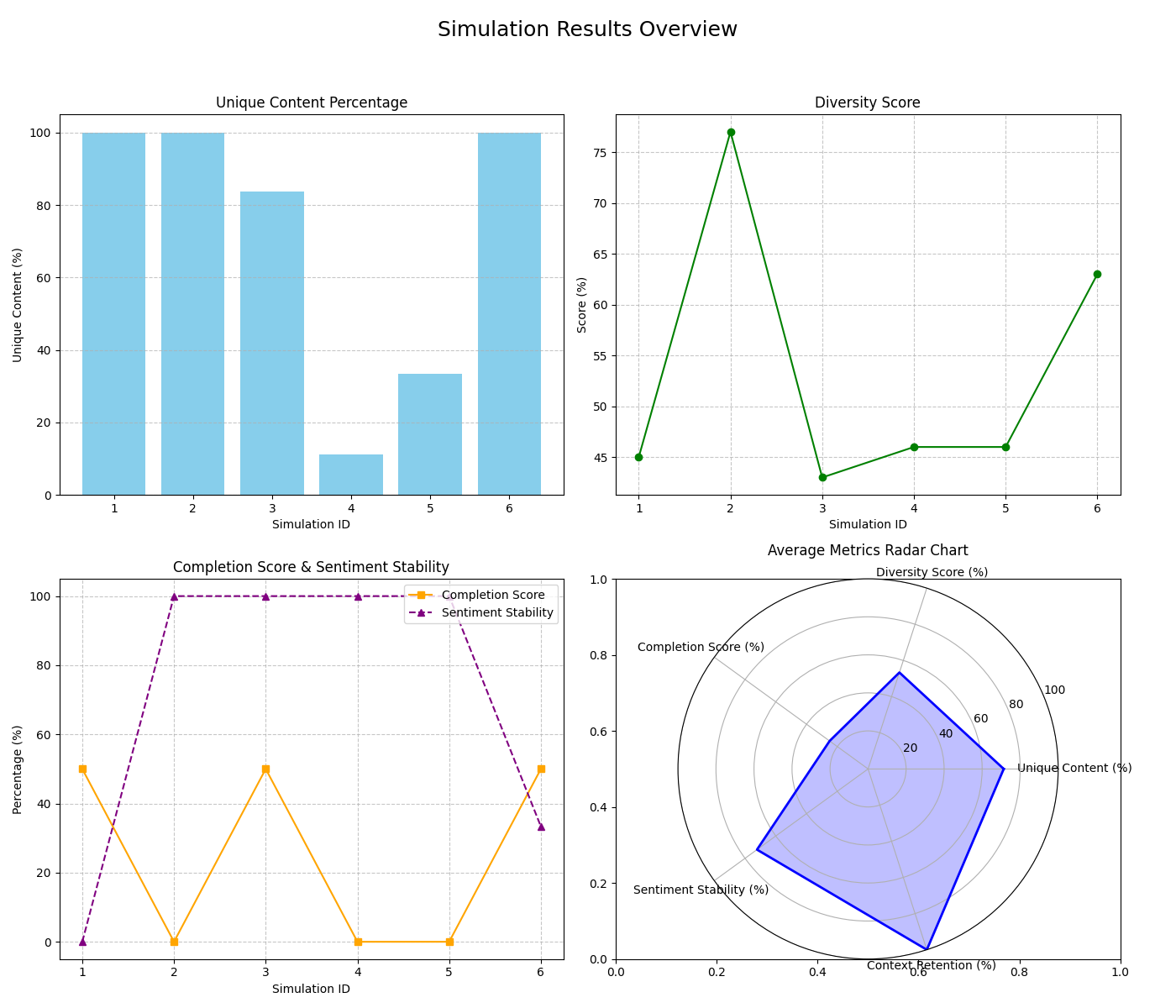}
\caption{Simulation results.}
\label{fig:performance_metrics}
\end{figure}

To provide a comprehensive overview, the graphical representation in Figure~\ref{fig:performance_metrics} illustrates key performance metrics derived from the simulations:

\begin{itemize}
    \item {{Unique content percentage:}
} A bar chart highlights the proportion of unique content generated in each simulation, reflecting the level of creative and non-redundant output.
    \item {{Diversity score:}} A line graph presents the diversity scores across simulations, emphasizing variations in the breadth and inclusivity of content.
    \item {{Completion score and sentiment stability:}} A combined plot showcases completion rates and sentiment stability trends, illustrating the balance between task execution and emotional consistency.
    \item {{Radar chart of average metrics:}} A radar chart summarizes the overall performance metrics, including unique content percentage, diversity score, completion score, context retention, and sentiment stability, offering an integrated view of agent performance.
\end{itemize}

\vspace{-12pt}

These visualizations offer an intuitive understanding of the simulation outcomes, enabling clear identification of strengths and areas for improvement. For instance, the unique content bar chart highlights simulations with high redundancy, while the radar chart underscores the overall balance among key performance indicators.

\subsection{Quality Measures and Performance Metrics}

{
To evaluate the effectiveness of the CogniSim framework, a set of quantitative and qualitative measures was used, capturing both the efficiency of cognitive agents and their adherence to best practices. These metrics included Task Completion Time, Quality of Deliverables, Communication Efficiency, Resource Utilization, Adaptability, and Compliance with Agile Practices.
Task Completion Time measures the duration taken by agents to complete assigned tasks, providing insights into their efficiency and speed~\cite{mazumder_dataperf_2023}. Quality of deliverables is assessed through code quality metrics such as cyclomatic complexity, code coverage, and adherence to coding standards~\cite{fenton_software_2015}. Communication efficiency is evaluated based on the frequency and clarity of interactions between agents and between agents and members of the human team, indicating the effectiveness of collaboration~\cite{chudziak2008communication}. Resource utilization monitors how effectively agents allocate and manage resources, reflecting optimization capabilities~\cite{qiao_taskweaver_2024}. Adaptability gauges agents' responsiveness to evolving project requirements and environmental conditions~\cite{weiss_multiagent_2013}, while compliance with Agile Practices ensures alignment with established Agile and SAFe methodologies, maintaining industry best practices~\cite{ebert_scaling_2017}.
}
{
These comprehensive metrics facilitate a multidimensional evaluation of agent performance. Continuous monitoring and analysis enable the identification of areas for improvement, guiding iterative enhancement of the CogniSim ecosystem. Furthermore, analyzing the impact of varying simulation parameters, such as model type, temperature settings, number of iterations, and agent roles, provides deeper insight into how these factors influence the metrics. For example, adjusting the temperature parameter affects the variability of agent responses and, in turn, influences Communication Efficiency and Quality of Deliverables. Increasing the number of iterations allows observation of agent behavior over extended interactions, offering further understanding of their Adaptability and Compliance with Agile Practices. By systematically refining these parameters, optimal configurations can be identified to enhance overall system performance.
}

\subsection{Key Findings and Implications}

The experiments reveal that cognitive agents can effectively handle structured Agile phases (e.g., PI Planning, Iteration Execution) by producing relevant user stories, code snippets aligned with predefined requirements, and infrastructure scripts for CI/CD pipelines. {For example, during Iteration Execution, agents delivered code integrating a payment API and generated corresponding test cases, thereby demonstrating the creation of tangible software artifacts.}

However, the agents’ effectiveness was mixed in more open-ended phases (e.g., Inspect and Adapt), highlighting challenges in adaptability and innovation without further prompt engineering. The choice of model type (GPT-4 vs. GPT-3.5-turbo) and parameter settings influenced the quality, clarity, and timeliness of outputs. While GPT-4 excelled in generating well-structured code and documentation, lower temperatures enhanced precision and~reliability.

{
In summary, these experiments answer our research questions (RQ1 and RQ2) by showing that cognitive agents can simulate various Agile roles effectively, produce meaningful software artifacts, and maintain performance quality under controlled experimental conditions. Adjusting independent variables (such as model type and iteration length) allowed us to identify settings that optimize dependent variables (such as code quality or responsiveness to changing requirements). Future work will focus on refining these experimental methods, incorporating more diverse project types, and validating results with real-world developer feedback.
}

\section{Future Work}
\label{sec:future_work}

The future directions of this study encompass multiple pivotal areas aimed at enhancing the integration of cognitive agents, Large Language Models, and Multi-Agent Systems within Agile development frameworks. These directions focus on research extensions, technological advancements, and the addressing of ethical concerns, ensuring the continuous evolution and practical applicability of the proposed framework.

\subsection{Research Extensions}

The integration of cognitive agents, Large Language Models, and Multi-Agent Systems within Agile development frameworks offers substantial opportunities for further research. A pivotal area for extending the current study involves exploring the scalability and interoperability of the LLM ecosystem across software projects of different sizes and complexities. This entails assessing the ecosystem's performance through pivotal metrics such as sprint completion times, defect rates, and adherence to project deadlines. By employing simulations and load testing, researchers can evaluate the system's efficiency and effectiveness as project sizes expand~\cite{horling_survey_2004, choinski_ontological_2009}.
\newpage
Another significant avenue for investigation is the enhancement of models for human--AI collaboration. Developing pioneering interfaces and mechanisms for mutual learning between human developers and AI agents can substantially enhance team dynamics. Refining MAS architectures to align more effectively with Agile practices may enable LLMs to simulate and facilitate human-like interactions within software development \mbox{teams~\cite{cabrero-daniel_ai_2023, guo_large_2024}}. This includes embedding predictive analytics into MAS to provide actionable insights into potential project delays, allowing teams to implement corrective measures~proactively.

Moreover, practical validation of the proposed ecosystem remains essential. Deploying the system across real-world software projects will facilitate the evaluation of its impact on productivity, quality, and team dynamics. Integration testing to ensure compatibility with tools such as Continuous Integration/Continuous Deployment (CI/CD) pipelines and issue trackers is crucial for its seamless incorporation into existing development workflows~\cite{choinski_ontological_2009}. Empirical studies in this context will generate valuable feedback for refining the system and addressing implementation challenges.

\subsection{Technological Advancements}

Emerging technologies present numerous opportunities to further enhance the proposed framework. Continuous advancements in AI and LLMs necessitate ongoing updates and retraining to sustain system effectiveness and adherence to ethical standards~\cite{tariverdi_trust_2024}. Investigating innovative MAS architectures capable of dynamically adapting to complex project requirements may lead to more resilient and flexible systems~\cite{guo_large_2024, lin_soen-101_2024}. Incorporating sophisticated natural language processing capabilities can enhance the contextual understanding and emotional intelligence of cognitive agents, enabling them to better interpret human inputs and respond effectively across diverse scenarios.

Advancements in cloud computing and distributed systems also offer robust support for the scalability and efficiency of MASs. By leveraging cloud services, such as Amazon Web Services, automation of tasks including load testing, cost management, and service health monitoring can be achieved, thereby improving the system's ability to manage tasks and mitigate risks~\cite{choinski_ontological_2009}. Additionally, integrating emerging technologies like blockchain for secure data management~\cite{dorri_multi-agent_2018} and the Internet of Things (IoT) for data collection and analysis can broaden the applicability and robustness of the framework.

By staying attuned to these technological advancements, the framework can evolve to address the changing demands of software engineering. Such evolution is crucial for maintaining relevance and effectiveness, ensuring that the system remains capable of addressing dynamic challenges in project management and software development.

\subsection{Security and Privacy Considerations}

As AI technologies become integral to project management, addressing ethical considerations is crucial, particularly regarding data privacy, security, and fairness. Protecting sensitive information requires robust encryption methods, stringent access controls, and adherence to data protection regulations such as GDPR. Transparency in AI-driven decision-making processes further fosters trust among team members and stakeholders. Bias mitigation in AI models is another critical challenge, addressed through continuous monitoring of outputs to identify and rectify biases caused by imbalanced training data or flawed algorithms. Employing fairness-aware machine learning techniques and promoting diversity in training datasets are effective strategies to reduce bias. Furthermore, fostering an organizational culture that prioritizes ethical AI usage through training and awareness programs supports the responsible and trustworthy integration of AI technologies~\cite{tariverdi_trust_2024}.
\newpage
Establishing clear guidelines and providing training for both human team members and AI agents can nurture a collaborative working environment. This includes stressing accountability, encouraging responsible AI usage, and facilitating open discussions about ethical dilemmas that may arise during project development~\cite{amirkhani_consensus_2022}.

{
However, LLMs may still inadvertently introduce biases or misinterpret nuanced project requirements, particularly when dealing with ambiguous instructions or culturally specific contexts. Their knowledge is derived from training data that can reflect historical imbalances or inaccuracies, potentially affecting the prioritization and communication of tasks. Recognizing these limitations is essential to ensure that human oversight and continuous refinement of the model remain central aspects of a reliable and equitable software project management ecosystem \cite{echterhoff-etal-2024-cognitive}.
}

\subsection{Integrating Cognitive Agents into the Enterprise-Wide Agile Scaling Framework}

\begin{figure}[H]

  \includegraphics[width=.99\columnwidth,keepaspectratio]{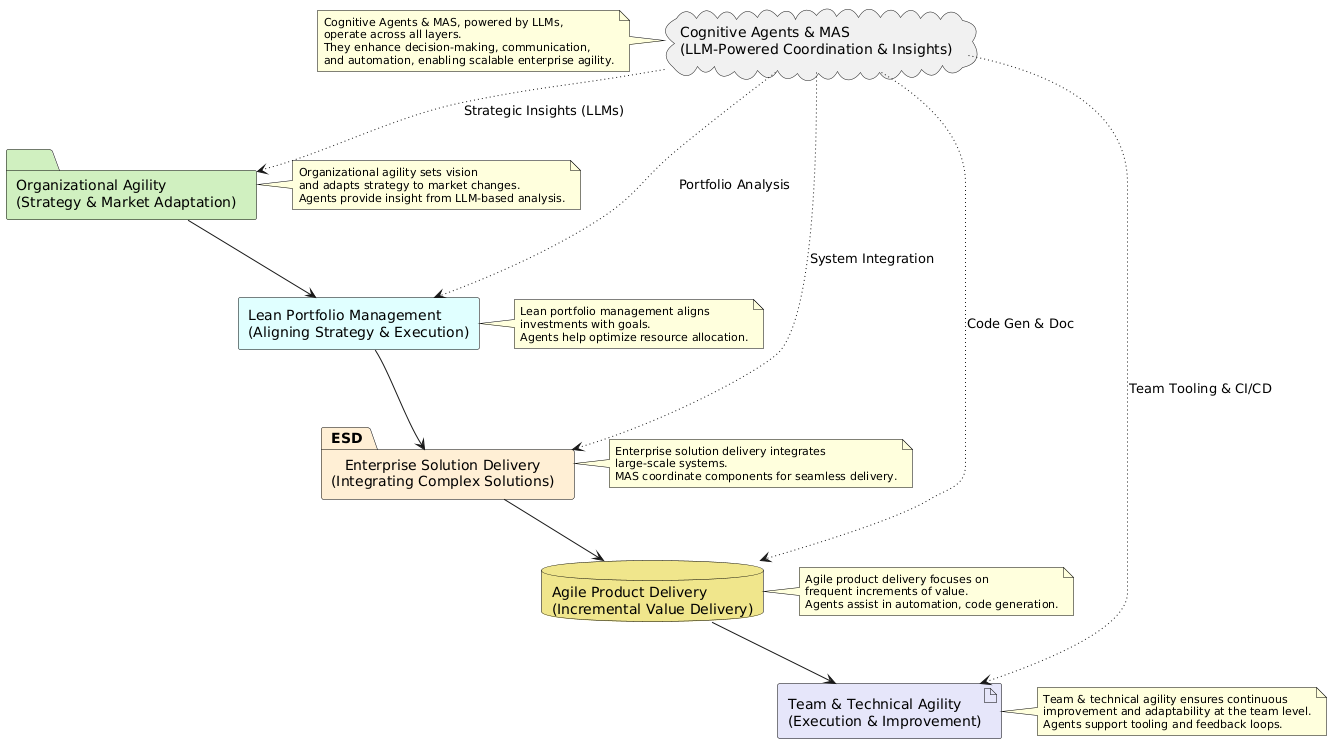}
  \caption{Conceptual enterprise-scale Agile layers with cognitive agents and MASs (inspired by principles in frameworks such as SAFe 6.0~\cite{scaledagile2024safescrum,safe6.0}).}
  \label{fig:conceptual_enterprise_layers}

\end{figure}

The Scaled Agile Framework is widely adopted for scaling Agile practices in large organizations, providing a structured approach through its core conceptual layers, which collectively enable organizations to achieve business agility~\cite{safe6.0}. As illustrated in Figure~\ref{fig:conceptual_enterprise_layers}, these layers emphasize the progression from organizational strategy down to team-level execution and highlight how cognitive agents and MASs can support each level.

By leveraging cognitive agents and Multi-Agent Systems at these higher conceptual layers, organizations can enhance decision-making, streamline operations, and foster continuous alignment of business strategies. For example, cognitive agents can automate data analysis to support strategic planning, ensuring that initiatives remain aligned with shifting market demands and organizational priorities. Additionally, MASs can optimize resource allocation by providing real-time insights, enabling more adaptive budgeting and portfolio management.

Future work involves exploring how the integration of cognitive agents and Large Language Models can further enhance enterprise-level Agile frameworks. By embedding cognitive agents into these layers, organizations can automate coordination tasks, improve communication efficiency, and facilitate decision-making processes. For instance, cognitive agents could assist in strategic alignment through predictive analytics, or in Agile Product Delivery by automating code generation and documentation~\cite{chen_evaluating_2021,svyatkovskiy_intellicode_2020}.

Moreover, incorporating Multi-Agent Systems powered by LLMs can support cross-team collaboration within scaled Agile initiatives. These agents can simulate human-like interactions, enabling more effective coordination across teams and enhancing the scalability of Agile practices~\cite{cruz_transforming_2024}.

Investigating the challenges and opportunities of integrating cognitive agents into enterprise-level Agile frameworks is a promising area for future research. This includes assessing the impact on team dynamics, addressing potential ethical concerns, and developing best practices for implementation.

\section{Conclusions and Summary}
\label{sec9}

The conclusions drawn from this study emphasize the transformative potential of integrating cognitive agents, Multi-Agent Systems, and Large Language Models into Agile development frameworks. By exploring the synergy between cognitive and autonomous systems, this research provides a comprehensive approach to enhancing project management processes and achieving improved productivity and collaboration. The findings underscore the importance of leveraging advanced technologies to address contemporary challenges in software engineering, ensuring adaptability and scalability in complex project~environments.

\subsection{Summary of Contributions}

This study has demonstrated the substantial advancements achieved through the integration of cognitive agents, Multi-Agent Systems, and Large Language Models into Agile development frameworks such as SAFe and SCRUM. {Rather than making definitive claims about complex task automation, decision-making improvements, or productivity gains, we now present these as preliminary high-level opportunities indicated by our initial simulations. For example, the framework showed potential in automating routine backlog item refinements and code snippet generation, offered preliminary indications of more autonomous architectural recommendations, and hinted at reduced turnaround times for certain project tasks. However, due to space and scope constraints, rigorous quantification of these gains or direct comparisons against human-led baselines are deferred to future~research.}

\begin{figure}[H]

  \includegraphics[width=0.6\columnwidth,keepaspectratio]{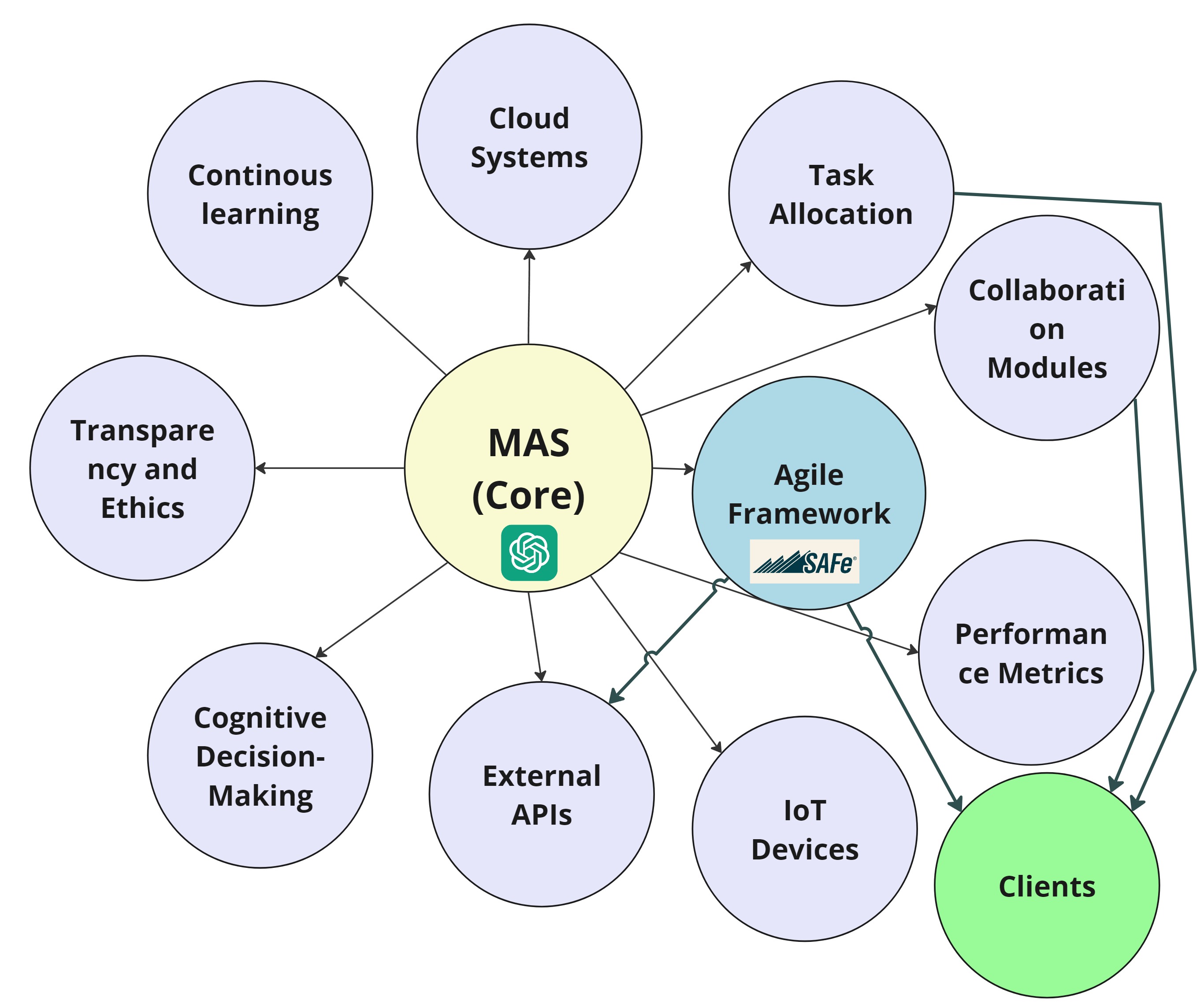}
  \caption{MAS concept diagram.}
  \label{fig:MASConceptDiagram}
\end{figure}

The research introduced a cohesive framework that facilitates improved task management and fosters collaboration among team members. The utilization of cognitive agents has proven effective in supporting continuous adaptation to evolving project needs, aligning seamlessly with the principles of Agile methodologies. Concept diagrams provided in Figure~\ref{fig:MASConceptDiagram} illustrate how the MAS interacts with critical project management components and external systems. These diagrams emphasize the system's ability to {theoretically} automate task allocation, problem-solving, collaboration, and documentation, while ensuring continuous learning, adaptability, and transparency within the system. {Specific empirical validation of these capabilities will be addressed in future studies.}

Additionally, \textls[+15]{the practical applications of MASs and LLMs extend beyond software development and project management into domains such as healthcare, education, and financial modeling. These systems streamline workflows, reduce the cognitive burden on human teams, and improve communication between developers, clients, and \mbox{stakeholders~\cite{amirkhani_consensus_2022, guo_large_2024}}. By addressing common Agile challenges, including communication gaps and scaling difficulties, the research provides valuable insights into improving Agile project~outcomes.}

\subsection{Implications for Practice}

The integration of these technologies presents significant implications for Agile software project management. Organizations that adopt MASs and LLMs can realize strategic benefits, including enhanced decision-making capabilities, improved risk management, and better project insights. Automating routine tasks enables human team members to focus on strategic project aspects, thereby boosting team productivity and fostering innovation.

Furthermore, the ability of these systems to handle advanced projects with greater precision and reduced operational costs can confer a competitive advantage in the software industry. By ensuring compatibility with existing cloud platforms and software libraries, the proposed architectures enable seamless incorporation into development practices~\cite{choinski_ontological_2009}. This adaptability is critical in software engineering, where flexibility and responsiveness to change are essential for success.

\subsection{Final Remarks}

In conclusion, the integration of cognitive agents, MASs, and LLMs into Agile development frameworks signifies a pivotal advancement in software engineering. While the potential benefits are considerable, it is essential to address the challenges associated with these technologies, particularly concerning data privacy, security, and transparency in AI-driven decisions~\cite{tariverdi_trust_2024}. Future research should prioritize refining human--AI collaboration models, enhancing agent capabilities, and ensuring that ethical considerations remain central to technological progress.

The ongoing evolution of these systems will undoubtedly shape the future of software engineering, enabling organizations to manage complex projects more effectively and efficiently. By embracing these innovations, the industry can anticipate improved decision-making processes, optimized workflows, and dynamic adaptation to project needs, driving progress and achieving success in an increasingly competitive landscape.

\vspace{6pt}
\authorcontributions{Conceptualization, K.C., J.A.C. and E.N.-S.; Methodology, K.C. and J.A.C.; Software, K.C.; Validation, K.C.; Formal analysis, J.A.C. and E.N.-S.; Investigation, K.C. and J.A.C.; Data curation, K.C.; Writing---original draft, K.C. and J.A.C.; Writing---review \& editing, E.N.-S.; Visualization, K.C.; Supervision, J.A.C.; Project administration, E.N.-S. All authors have read and agreed to the published version of the manuscript. 
}

\funding{{ This research received no external funding } 
}

\dataavailability{{The} original contributions presented in this study are included in the article. Further inquiries can be directed to the corresponding author. 
} 

\conflictsofinterest{{ The authors declare no conflict of interest.} 
}

\begin{adjustwidth}{-\extralength}{0cm}

\reftitle{References}

\PublishersNote{}

\end{adjustwidth}

\end{document}